\documentclass[10pt,twocolumn]{IEEEtran}
\usepackage{amsmath}
\usepackage{amsfonts}
\usepackage{amssymb}
\usepackage{dsfont}
\usepackage{enumerate}
\usepackage{multirow}
\usepackage{color}
\usepackage{graphicx}
\usepackage{epstopdf}
\usepackage{caption}
\usepackage{cases}
\usepackage{subfigure}
\usepackage{cite}
\usepackage{bm}
\usepackage{algorithm}
\usepackage{algorithmicx}
\usepackage{algpseudocode}
\usepackage{setspace}
\usepackage{threeparttable}
\newtheorem{definition}{Definition}

\newtheorem{lemma}{Lemma}
\newtheorem{remark}{Remark}

\newtheorem{proposition}{Proposition}
\allowdisplaybreaks[4]

\newcommand{\ri}{\rightarrow\infty}

\newcommand{\e}{\mathrm{e}}
\newcommand{\rmd}{\mathrm{d}}
\newcommand{\deq}{\triangleq}

\begin{document}
\title{Jamming Aided Covert Communication  with Multiple Receivers}
\author{Ke-Wen Huang, Hao Deng, Hui-Ming Wang, \emph{Senior Member, IEEE}
\thanks{\scriptsize
K.-W. Huang and H.-M. Wang are with the School of Information and Communications Engineering, and also with the Ministry of Education Key Lab for Intelligent Networks and Network Security, Xi'an Jiaotong University, Xi'an 710049, Shaanxi, P. R. China
(e-mail: {\tt
xjtu-huangkw@outlook.com, xjbswhm@gmail.com}).
}
\thanks{\scriptsize
H. Deng is with the School of Physics and Electronics, Henan University, Kaifeng 475001, China (e-mail: {\tt gavind@163.com}).
}

}
\IEEEtitleabstractindextext{
\begin{abstract}
We consider that a transmitter covertly communicates with multiple receivers under the help of a friendly jammer. The messages intended for different receivers are transmitted in mutually orthogonal frequency bands. An adversary observes all these frequency bands aiming at detecting whether or not communication occurs, while the friendly jammer broadcasts jamming signals to degrade the detection performance of the adversary.
We consider a block Rayleigh fading channel model and evaluate the performance of covert communication  in two situations:
1) the wireless channels vary slowly such that the transmission ends within one channel coherent time block, and
2) the wireless channels vary fast such that the wireless channels have changed several times before the whole transmission is finished.
In the former case, subject to a covertness constraint,
we maximize the sum of the effective rates by optimizing the transmit power allocation and the transmission rate for each receiver.
In the latter case, we take the channel training process into consideration,
and subject to a covertness constraint,  we maximize the sum of the ergodic rates by optimizing the power allocation and the pilot length.
Though both of the two optimization problems are non-convex, we presented methods to find their global optimal solutions.
Besides, we also present methods to find sub-optimal solutions with lower computational complexities.
Numerical results are presented to evaluate the performance under the two situations.
\end{abstract}

}

\maketitle

\IEEEdisplaynontitleabstractindextext
\IEEEpeerreviewmaketitle
\section{Introduction}

Providing endogenous network security represents one of the new paradigm shifts of next generation wireless networks \cite{X.You,B.AiPoIEEE2020}.
Recently, covert communication has gained considerable attention for its ability to hide the occurrence of the communication itself  \cite{B.A.BashCM2015,S.Yan219Arxiv}.
Consider a wireless system wherein a transmitter sends its message to an intended receiver.
An adversary listens to the wireless channels aiming at detecting whether or not transmission occurs, posing a threat on system security.
The technique of covert communication enables the transmitter to reliably communicate with the receiver while ensuring a high probability that the detector of the adversary produces an incorrect result, which greatly enhances wireless security.

The information-theoretic performance limits of covert communication have been studied in \cite{B.A.BashJSAC2013,P.H.CheISIT2013,M.R.BlochTIC2016,L.WangTIT2016}.
It was revealed in \cite{B.A.BashJSAC2013} that covert communication over additive white Gaussian noise  (AWGN) channels is subject to the so called \emph{square root law} (SRL). Specifically, a transmitter is able to reliably and covertly transmit at most $O(\sqrt{n})$ bits to a receiver in $n$ channel uses as $n\ri$. \cite{P.H.CheISIT2013,M.R.BlochTIC2016,L.WangTIT2016} further extended the SRL to binary symmetric channels and general discrete memoryless channels.
The SRL indicates that the covert communication rate is asymptotically zero, i.e., $\lim_{n\rightarrow \infty }\frac{\sqrt{n}}{n}  = 0$. Some recently efforts have been devoted to improving the covert communication performance.
It has been revealed that by exploiting the adversary's uncertainty on the statistical information of its channel outputs, the covert communication performance can be greatly improved, and in some cases, non-vanishing covert communication rates exist, for example, when the adversary has uncertainty on its noise power \cite{D.Goeckel2016CL,S.LeeJSTSP2015} or when the adversary is uncertain about the transmission time \cite{B.A.BashTWC2016,M.R.BlochITW2016,K.WHuangTCOM2020}.

\subsection{Wireless covert communication}
Covert communication over practical wireless channels has also been extensively investigated.
Due to the openness of wireless media and the randomness of wireless environment,
wireless transmissions inevitably suffer from co-channel interference.
Though co-channel interference is harmful to normal wireless communication process, it also leads to a poor detection performance at the adversary, and thus its impacts have been studied in many existing works \cite{B.He2018TWC,T.-X.Zheng2019TWC,
Y.Jiang2020TVT,J.Hu2019TWC,J.Hu2018TWC,D.Goeckel2018SPAWC,K.Shahzad2018TWC,F.Shu2019WCL,J.Hu2019TVT,
T.V.Sobers2017TWC,R.Soltani2018TWC,K.Li2020TWC,M.Forouzesh2020TVT,O.ShmuelISIT2019}.

Covert communication under co-channel interference has been studied in \cite{B.He2018TWC,T.-X.Zheng2019TWC,
Y.Jiang2020TVT,J.Hu2019TWC,J.Hu2018TWC,D.Goeckel2018SPAWC}.
In \cite{B.He2018TWC,T.-X.Zheng2019TWC}, the locations of the interferers were modeled as a Possion point process. Subject to  a covert outage probability upper bound, the covert throughput was maximized by optimizing the transmit power and the transmission rate.
Covert communication in device-to-device (D2D) underlaying cellular networks was studied in \cite{Y.Jiang2020TVT}, where the wireless signals of the cellular users were treated as interference and used to hide the communications between D2D pairs.
\cite{J.Hu2019TWC,J.Hu2018TWC} studied covert communication in one-way relay networks, and the authors maximized the effective covert rate achieved by the relay.
In \cite{D.Goeckel2018SPAWC}, the dynamicity of the interference environment was considered, and the authors studied the covert throughput scaling law with respect to the codeword length and the variation rate of the background interference environment.

The performance of covert communication can be improved by letting a friendly jammer deliberately broadcast jamming signal to degrade the detection performance of the adversary \cite{T.V.Sobers2017TWC,R.Soltani2018TWC,
		K.Li2020TWC,M.Forouzesh2020TVT,O.ShmuelISIT2019}.
Remarkably, \cite{T.V.Sobers2017TWC} theoretically showed the existence of constant covert communication rates provided that the  adversary does not know the jamming power or the instantaneous realization of the jamming channel.
In \cite{K.Li2020TWC}, a truncated channel inversion power adaption scheme was proved to be optimal in term of minimizing the outage probability subject to a covertness constraint.
In \cite{R.Soltani2018TWC}, the covert throughput in a random network was studied, wherein the jammer who is closest to the adversary broadcasts jamming signal.
In \cite{M.Forouzesh2020TVT} and \cite{O.ShmuelISIT2019}, a friendly jammer was assumed to have multiple antennas and use beamforming  to maximize its ability to degrade the detection performance of the adversary.

Instead of relying on external friendly jammers, the works in \cite{K.Shahzad2018TWC,F.Shu2019WCL,J.Hu2019TVT} assumed that the receiver operates in full-duplex mode, i.e., simultaneously receiving the signal from the transmitter and broadcasting jamming signal to the deteriorate detection performance of the adversary.
Due to the imperfect self interference cancellation, the trade-offs between the covertness and the
reliability should be carefully designed. In particular, \cite{K.Shahzad2018TWC} and \cite{F.Shu2019WCL} designed the transmit power of the transmitter and the jamming power of the receiver to maximize the detection error probability at the adversary subject to a lower bound on the effective throughput, and to minimize the outage probability subject to a lower bound on the detection error probability at the adversary, respectively. In \cite{J.Hu2019TVT}, an on-off transmission strategy were proposed to optimize the covert communication performance.

\subsection{Motivations, challenges, and contributions}
\label{MotivationSec}
Existing works have presented important insights on the achievable performance of covert communication over wireless channels, however, most of them have only discussed the transmissions from a single transmitter to a single receiver.
Motivated by this observation, in this paper, we study a covert communication scenario wherein a transmitter simultaneously communicates with multiple receivers over wireless fading channels.
Under the help of a friendly jammer, the transmitter communicates with different receivers in different and mutually orthogonal frequency bands (namely the system works in a frequency-division multiplexing manner). The adversary is able to observe all the frequency bands to make a decision on whether communication occurs or not, whereas we let the friendly jammer broadcast jamming signals in all these frequency bands to degrade the detection performance of the adversary.

It is worth noting that compared to the single-receiver case investigated in literature, analyzing the covertness of the communication becomes more challenging in the considered multiple-receiver case.
Specifically, in existing works such as \cite{B.He2018TWC,T.-X.Zheng2019TWC,
Y.Jiang2020TVT,J.Hu2019TWC,J.Hu2018TWC,K.Shahzad2018TWC,J.Hu2019TVT,R.Soltani2018TWC,K.Li2020TWC,
M.Forouzesh2020TVT,O.ShmuelISIT2019}, the optimal detector of the adversary was shown to be an energy-based detector \cite{T.V.Sobers2017TWC}, i.e., comparing the received energy to a predesigned threshold. By exploiting the simple mathematical form of the energy-based detector, the optimal detection performance of the adversary can be accurately characterized by analyzing the false alarm and missed detection probabilities of the optimal energy-based detector.
However, in our case, due to the fact that the fading channel coefficients in different frequency bands are different, the energy-based detector becomes strictly suboptimal, which prevents us from analyzing the covertness of the communication by using the methods adopted in existing works.

In addition, the transmit power allocation problem is also a key issue in the considered multiple-receiver scenario.
Since the adversary is able to observe all the frequency bands, increasing the transmit power in each single frequency band will lead to an improved detection performance at the side of the adversary. Then, a natural question is that subject to a certain constraint on the covertness, how to allocate the transmit power to the multiple receivers in different frequency bands in order to achieve the optimal communication performance. Based on this observation, the issue of power allocation at the side of the transmitter constitutes the main problem that will be studied in this paper.

The contributions of this paper are summarized as follows,
\begin{enumerate}
\item We consider that a transmitter covertly communicates with multiple receivers in mutually orthogonal frequency bands under the help of a jammer. Depending on the temporal dynamic properties of the wireless channels, we consider two different situations:
    \emph{1)} the wireless channels vary slowly over time and a single transmission of the transmitter terminates before the wireless channels have changed,
    and
    \emph{2)} the wireless channels vary fast and during a single transmission period, the wireless channels change several times.
    For convenience, the wireless channels are said to be \emph{quasi-static} and \emph{fast-varying} in these two situations, respectively. For both situations, analytically tractable upper bounds on the total variation distance  between  $f_1$ and $f_0$
    are derived to characterize the covertness of the communication,
    where $f_1$ and $f_0$ denote the probability density functions (PDFs) of the adversary's channel outputs given that covert communication occurs and does not occur, respectively.
\item Under the condition that the channels are quasi-static, we maximize the sum of the effective communication rates of the multiple receivers subject to a constraint on covertness by jointly optimizing the power allocation and the transmission rate for each receiver. The established optimization problem is non-convex, however, by exploiting its monotonic properties, we can obtain the global optimum by using the polyblock outer approximation (POA) method.
    Besides, we also present a computationally efficient method based on successive convex approximation (SCA) to search for a sub-optimal solution.
\item Under the condition that the channels are fast-varying, due to the limited channel coherent time, we take the channel training process into consideration. Accordingly, we maximize the sum of the ergodic rates subject to a constraint on covertness by jointly optimizing the pilot length  and the power allocation.
    An exhaustive search (ES) based method is proposed to calculate the global optimal solution.
    In addition, we also present an alternating optimization based method to obtain a sub-optimal solution, which has a lower computational complexity than the ES based method but causes little performance loss.
\end{enumerate}

The rest of this paper is organized as follows:
in Section II, we introduce the system model;
in Section III and Section IV, we study the covert communication performance under the conditions that the channels are quasi-static and fast-varying, respectively;
Numerical results are presented in Section V;
and finally, Section VI concludes the paper.

\emph{Notations:} $\mathcal{C}$, $\mathcal{R}$, and $\mathcal{R}_+$ denote the set of complex, real, and non-negative numbers, respectively.
$(\cdot)^T$ and $(\cdot)^H$ denote transpose  and conjugate transpose, respectively.
$\mathbb{E}(\cdot)$ and $\mathbb{P}(\cdot)$ denote mathematical expectation and probability, respectively.
$|\cdot|$ and $\|\cdot\|$ denote the absolute value and the norm, respectively.
$\mathcal{CN}(\cdot,\cdot)$ and $\mathcal{E}(\cdot)$ denote the complex Gaussian and exponential distributions, respectively.
Diagonal matrix is denoted by $\mathrm{diag}(\cdot)$.
For any two real vectors $\bm{x},\bm{y}\in \mathcal{R}^n$, $\bm{x}\geq \bm{y}$ if $x_i \geq y_i$ for $\forall i=1,2,\cdots,n$
where $x_i$ and $y_i$ are the $i$-th elements of $\bm{x}$ and $\bm{y}$, respectively.
$\bm{I}_m$ denotes the $m$-by-$m$ identity matrix.
$f(x) = O(g(x))$ means that $\lim_{x\rightarrow\infty} \frac{f(x)}{g(x)}\leq c$ for some constant $c>0$. $f(x) = o_x(1)$ means that $\lim_{x\rightarrow \infty} f(x) = 0$.
For any two PDFs $f_1(\bm{x})$ and $f_0(\bm{x})$ defined in $\mathcal{S} \subseteq \mathcal{R}^n$ (or $\mathcal{C}^n$), the total variation distance between $f_1(\bm{x})$ and $f_0(\bm{x})$ is defined as  $\mathbb{V}(f_1,f_0) = \frac{1}{2}\int_{\mathcal{S}} |f_1(\bm{x})- f_0(\bm{x})| \rmd \bm{x}$, and
the Kullback-Leibler (KL) divergence is defined as $\mathbb{D}(f_1||f_0) = \int_{\mathcal{S}} f_1(\bm{x})\ln\frac{f_1(\bm{x})}{f_0(\bm{x})} \rmd \bm{x}$.

\section{System Model}
\begin{figure}[t]
  \centering
  \includegraphics[width=3  in]{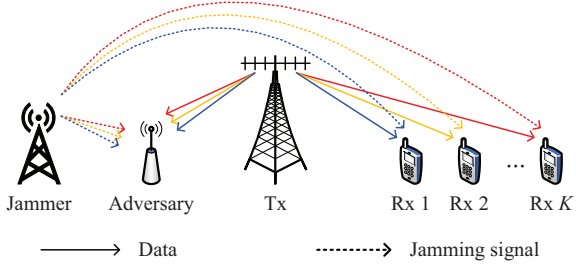}\\
  \caption{A comprehensive system model, wherein Tx represents the transmitter and Rx $k$ is the $k$-th receiver.}
  \label{SM}
  \vspace{-6mm}
\end{figure}

Consider that a transmitter communicates with $K$ receivers in $K$ mutually orthogonal frequency bands ($K\geq 1$).
An adversary aims to detect the occurrence of the communication by observing the $K$ frequency bands.
The goal of the transmitter is to reliably communicates with the $K$ receivers while ensuring that the adversary is unable to effectively detect the existence of the communication.
Following the works in \cite{T.V.Sobers2017TWC,R.Soltani2018TWC,K.Li2020TWC,M.Forouzesh2020TVT,O.ShmuelISIT2019}, we consider that a friendly jammer broadcasts jamming signals in order to deteriorate the detection performance of the adversary, which potentially helps the transmitter to accomplish its transmission covertly.
A comprehensive system model is depicted in Fig. \ref{SM}.

\subsection{Signal model and basic assumption}
We assume that the transmitter has $M$ ($M\geq 1$) antennas and that the $K$ receivers, the adversary, and the jammer each have a single antenna. For convenience, assume that the transmitter uses the $k$-th orthogonal frequency band to communicate with the $k$-th receiver where $1\leq k\leq K$.
We consider a block fading channel model. Specifically, the channel coefficients remain unchanged in each channel coherent time block and are identically and independently distributed (i.i.d.) in different time blocks.
Besides, the channel coefficients in different frequency bands are also i.i.d.
For simplicity, we assume that all the channels share the same channel coherent time.
In the $i$-th ($i\geq 1$) time block,
denote by $\bm{h}_{k,i}\sim\mathcal{CN}(\bm{0},\bm{I}_M)$ and  $\phi_{k,i}\sim \mathcal{CN}(0,1)$
the small-scale fading channels from the $k$-th receiver to the transmitter and the jammer, respectively.
Denote by $\bm{g}_{k,i}\sim\mathcal{CN}(\bm{0},\bm{I}_M)$ and $\varphi_{k,i}\sim \mathcal{CN}(0,1)$
the small-scale fading channels from the transmitter and the jammer to the adversary in the $k$-th frequency band, respectively.

We consider that the communication system works in time division duplex mode. In each channel coherent time block, which consists of $N$ ($N>1$) symbol periods, each receiver first sends a pilot sequence of length $N_t$ to the transmitter, and the transmitter estimates the channels by exploiting the channel reciprocity.
After that, the transmitter uses the remaining $N_d = N - N_t$ symbol periods to transmit its data to the receivers.
In the $i$-th time block, the pilot sequence received by the transmitter in the $k$-th frequency band is
\begin{align}
\label{ReversePilot}
\bm{Y}_{k,i} = \sqrt{P_{{\rm R}_k} S_{{\rm R}_k,{\rm T}}} \bm{h}_{k,i} \left(\bm{x}_{k}^{(p)}\right)^H + \bm{Z}_{k,i}, \quad 1\leq k\leq K,~ i\geq 1,
\end{align}
where $P_{{\rm R}_k}$ is the transmit power of the $k$-th receiver,
$S_{{\rm R}_k,{\rm T}}$ is the distance-based path loss between the $k$-th receiver and the transmitter,
$\bm{x}_{k}^{(p)} \in \mathcal{C}^{N_t \times 1}$ is the pilot sequence sent by the $k$-th receiver, and
$\bm{Z}_{k,i}\in\mathcal{C}^{M\times N_t}$ is the AWGN with each of its elements distributed as $\mathcal{CN}( 0 , \sigma_{k,\rm{T}}^2)$. For notation simplicity, we assume that
$P_{{\rm R}_1} = \cdots =P_{{\rm R}_K}=P_{\rm R}$ and
$\sigma_{1,\rm{T}}^2 = \cdots = \sigma_{K,\rm{T}}^2 = \sigma_{\rm{T}}^2$.
Based on \eqref{ReversePilot},  the minimum mean square error estimation of $\bm{h}_{k,i}$, denoted by $\hat{\bm{h}}_{k,i}$, and the estimation error, denoted by $\tilde{\bm{h}}_{k,i}$, are respectively given by
\begin{align}
\hat{\bm{h}}_{k,i}
&= \frac{\sqrt{P_{\rm R} S_{{\rm R}_k,{\rm T}}} }{N_t P_{\rm R} S_{{\rm R}_k,{\rm T}} + \sigma_{\rm{T}}^2} \bm{Y}_k\bm{s} \sim\mathcal{CN}\left(\bm{0}, \frac{N_t }{N_t  + \mu_k }\bm{I}_M \right), \label{EstimatedChannel} \\
\tilde{\bm{h}}_{k,i} &= \bm{h}_{k,i} - \hat{\bm{h}}_{k,i}
\sim\mathcal{CN}\left(\bm{0}, \frac{\mu_k}{N_t  + \mu_k}\bm{I}_M\right),\label{EstimatedChannelError}
\end{align}
where $\mu_k \triangleq \frac{\sigma_{ \rm{T}}^2}{P_{\rm R} S_{{\rm R}_k,{\rm T}}}$.
After obtaining $\hat{\bm{h}}_{k,i}$,  the transmitter adopts the maximum ratio transmission beamforming scheme to transmits its data to the $k$-th receiver, and the beamforming vector is given by
$\bm{b}_{k,i} = \frac{\hat{\bm{h}}_{k,i}} {||\hat{\bm{h}}_{k,i}||}$.
Accordingly,  the $k$-th receiver receives
\begin{align}
\bm{y}_{k,i} &=  \sqrt{P_{k} S_{{\rm R}_k,{\rm T}} }
\bm{h}_{k,i}^H \bm{b}_{k,i}\bm{x}_{k,i}^{(d)}  \nonumber \\
&\quad + \sqrt{Q_{k} S_{{\rm R}_k,{\rm J}} } \phi_{k,i} \bm{v}_{k,i}  + \bm{z}_{k,i},\label{Signalkl}
\end{align}
where $P_{k}$ is the transmit power allocated for the $k$-th receiver,
$\bm{x}_{k,i}^{(d)}\sim\mathcal{CN}(0,\bm{I}_{N_d})$ is the data sequence,
$Q_{k}$ is the jamming power that the jammer emits in the $k$-th frequency band,
$S_{{\rm R}_k,\rm J}$ is the path loss between the $k$-th receiver and the jammer,
$\bm{v}_{k,i}\sim\mathcal{CN}(0,\bm{I}_{N_d})$ is a sequence of the jamming signals transmitted by the jammer,
and  $\bm{z}_{k,i}\sim\mathcal{CN}(0,\sigma_{{\rm R}_k}^2\bm{I}_{N_d})$ is the AWGN.
It is worth noting that by \eqref{ReversePilot} and \eqref{Signalkl}, we assume that the jammer keeps silent during the channel training process and jams the $K$ frequency bands during the data transmission process.

\begin{remark}
In this paper, we aim to keep the adversary unaware of the transmission from the transmitter while the receivers are allowed to broadcast wireless signals.
Similar assumptions have also been adopted in existing works such as \cite{K.Shahzad2018TWC,F.Shu2019WCL,J.Hu2019TVT} wherein full-duplex receivers are exploited to broadcast jamming signals to improve covert communication performance. In our situation, we let the receivers broadcast pilot sequences so that the transmitter can estimate the wireless channels and design beamforming vectors.
\end{remark}

\subsection{Adversary model and metric for covertness}
From the perspective of the adversary, it does not know whether communication occurs or not, and the signals received by the adversary in the $i$-th time block can be modeled as
\begin{align}
\label{AdversarySignalModel}
\bm{w}_{k}(i) &= \left\{\begin{aligned}
&\sqrt{P_{k} S_{\rm A, T}} \bm{g}_{k,i}^H \bm{b}_{k,i} \bm{x}_{k,i}^{(d)} \\
&\quad\quad + \sqrt{Q_k S_{\rm A,J}} \varphi_{k,i} \bm{v}_{k,i}  + \tilde{\bm{z}}_{k,i} ,&& \mathcal{H}_{1},\\
&\sqrt{Q_k S_{\rm A,J}} \varphi_{k,i} \bm{v}_{k,i}  + \tilde{\bm{z}}_{k,i},&&\mathcal{H}_{0},
\end{aligned}\right.
\end{align}
where $k\in \{1,2,\cdots,K\}$, $\mathcal{H}_{1}$ and $\mathcal{H}_{0}$ stand for the hypotheses that communication occurs and does not occur, respectively, $\bm{w}_{k}(i)\in\mathcal{C}^{N_d\times1}$ is the signal received in the $k$-th frequency band,
$S_{{\rm A}, {\rm T}}$ and $S_{\rm A,J}$ are the path losses from the transmitter and the jammer to the adversary, respectively,
and $\tilde{\bm{z}}_{k,i}\sim\mathcal{CN}\left(\bm{0}, \sigma_{{\rm A}_k}^2\bm{I}_M\right)$ is the AWGN.
We assume that the adversary does not know the instantaneous channel realizations, i.e., $\varphi_{k,i}$ and $\bm{g}_{k,i}$ for $1\leq k\leq K$, but knows their statistic distributions. Besides, we assume that the jamming powers, i.e.,  $Q_k$ for $1\leq k\leq K$, are fixed constants.

The goal of the adversary is to determine whether or not communication occurs based on its received signals in the $K$ frequency bands.
We consider the following two scenarios depending on the temporal dynamic properties of the wireless channels:
\subsubsection{The channels are quasi-static} The channel coherent time block is sufficiently long, and the transmission from the transmitter terminates before the end of a channel coherent time block. In this case, we assume that the adversary makes a decision to determine whether communication occurs at the end of each time block.
\subsubsection{The channels are fast-varying} The wireless channels vary fast as compared to the length of a single transmission. Specifically, we consider that a single transmission is comprised of $L>1$ consecutive short channel coherent time blocks.
In this case, we assume that the adversary collects the signals received in the $L$ consecutive time blocks together to make a decision.

For a given detector of the adversary, denoted by $d$, let $\mathbb{P}_{\rm FA}(d)$ and $\mathbb{P}_{\rm MD}(d)$ be the false alarm and the missed detection probabilities, respectively.
In this paper, the communication between the transmitter and the multiple receivers is said to be \emph{$(1 - \epsilon)$-covert} if it satisfies that
\begin{align}
\min_{d} ~\mathbb{P}_{\rm FA}(d) + \mathbb{P}_{\rm MD}(d) \geq 1 - \epsilon, \label{CovertConstraint}
\end{align}
where $\epsilon\in(0,1)$ is a pre-fixed number.
Note that the minimization operation in \eqref{CovertConstraint} is with respect to the detector adopted by the adversary. This means that the covertness is guaranteed no matter how the adversary performs the detection \cite{B.A.BashJSAC2013}.
For convenience, we refer to \eqref{CovertConstraint} as the \emph{covertness constraint} in the subsequent part of this paper.

\begin{remark}
\label{OptimalAdversary}
From the viewpoint of the adversary, it solves a binary hypothesis testing problem which tests $\mathcal{H}_1$ against $\mathcal{H}_0$ defined in \eqref{AdversarySignalModel}. By \cite[Theorem 13.1.1]{E.Lehmann}, the optimal detector that minimizes the sum of the false alarm and the missed detection probabilities is the likelihood ratio test, which makes a decision by comparing the likelihood ratio with a pre-designed threshold. We assume that the adversary knows the exact values of $\hat{p}_k = P_kS_{\rm A, \rm T}$, $\hat{q}_k = Q_kS_{\rm A, \rm J}$, and $\sigma_{{\rm A}_k}^2$, $1\leq k\leq K$, so that it can perform the optimal detection.
It is also worth noting that the optimal detector is not equivalent to the energy-based detector due to the different fading channel coefficients in different frequency bands and time blocks, see e.g., \cite{D.Goeckel2018SPAWC,T.V.Sobers2017TWC}.
\end{remark}

\subsection{Covert communication problem formulation}
\subsubsection{Covert communication over quasi-static channels}
In this situation, we assume that the pilot sequence is sufficiently long so that the transmitter can accurately estimate the channel coefficients. As a result, according to \eqref{Signalkl}, the signal-to-noise ratio (SNR) of the $k$-th user is
\begin{align}
{\rm SNR}_{k} = \frac{ P_{k} S_{{\rm R}_k, {\rm T} } ||\bm{h}_{k}||^2}{ Q_{k} S_{{\rm R}_k,{\rm J}} \left| \phi_{k}\right|^2  + \sigma_{{\rm R}_k}^2}.
\end{align}
Here, we have omitted the time block index $i$ because in this case, the transmission terminates within a single time block.
Suppose that the transmission rate for the $k$-th receiver is $r_k > 0$.
Due the randomness of the interfering channel  $\phi_{k}$, an outage event occurs if $\ln(1 + {\rm SNR}_{k}) \leq r_k$, and the outage probability is given by
\begin{align}
\label{OutageProbability}
\mathcal{O}_{k} &\triangleq \mathbb{P}\left\{  {\rm SNR}_k  \leq \e^{r_k} - 1 \right\} \nonumber \\
&= \mathbb{I}\left\{ \e^{r_k} - 1 \leq \frac{ P_{k} S_{{\rm R}_k, {\rm T}} ||\bm{h}_{k}||^2}{ \sigma_{{\rm R}_k}^2} \right\}\nonumber \\
&\quad \times \e^{ - \frac{\frac{ P_{k} S_{{\rm R}_k,{\rm T}} ||\bm{h}_{k}||^2}{\e^{r_k} - 1} - \sigma_{{\rm R}_k}^2}{Q_{k} S_{{\rm R}_k,{\rm J}}} }.
\end{align}
We define the effective rate of the $k$-th receiver as
$r_k^{(e)} = (1 - \mathcal{O}_{k}) \times r_k$.
The goal of the transmitter is to maximize the sum of the effective rates of the $K$ receivers while ensuring that its transmission is at least $(1 - \epsilon)$-covert for some pre-fixed $\epsilon\in(0,1)$, i.e.
\begin{align}
\label{Opt1}
\max_{\bm{p} ,\bm{r}\in\mathcal{R}_+^K}&\quad r^{(e)} \triangleq \sum_{k=1}^K r_k^{(e)},\quad
{\rm s.t.} \quad\eqref{CovertConstraint},
\end{align}
where $\bm{p} = [P_1,P_2,\cdots,P_k]^T$ and $\bm{r} = [r_1,r_2,\cdots,r_K]^T$.

\subsubsection{Covert communication over fast-varying channels}
In this case, the wireless channel changes several times before the transmission is finished.
Without loss of generality, we assume that the transmission starts in the $1$-st time block and ends at the $L$-th time block.
Due to the limited channel coherent time, the pilot length, $N_t$, cannot be neglected as compared to the block length $N$. As a result, the channel estimation error becomes non-negligible.
To evaluate the communication performance, we reformulated \eqref{Signalkl} as follow,
\begin{align}
\bm{y}_{k,i} &=  \sqrt{P_{k} S_{{\rm R}_k,{\rm T}} } \mathbb{E}\{ \bm{h}_{k,i}^H \bm{b}_{k,i} \} \bm{x}_{k,i}^{(d)}\nonumber \\
&\quad + \sqrt{P_{k} S_{{\rm R}_k,{\rm T}}} ( \bm{h}_{k,i}^H \bm{b}_{k,i} - \mathbb{E}\{ \bm{h}_{k,i}^H \bm{b}_{k,i} \})\bm{x}_{k,i}^{(d)} \nonumber\\
&\quad + \sqrt{Q_k S_{{\rm R}_k,{\rm J}}} \phi_{k,i} \bm{v}_{k,i}  + \bm{z}_{k,i},
\end{align}
where $k\in \{1,2,\cdots,K\}$ and $i\in \{1,2,\cdots,L\}$. Following \cite[Theorem 1]{J.Jose2011TWC}, an achievable ergodic rate of the $k$-th receiver is
\begin{align}
\label{AchievableErgodicRate}
\bar{r}_k &= \frac{N - N_t}{N} \nonumber\\
&\times \ln\left( 1 + \frac{ P_{k} S_{{\rm R}_k,{\rm T}} | \mathbb{E}\{ \bm{h}_{k,i}^H \bm{b}_{k,i} \} |^2  }{ Q_k S_{{\rm R}_k,{\rm J}}   + P_{k} S_{{\rm R}_k,{\rm T}} \mathrm{Var}(\bm{h}_{k,i}^H \bm{b}_{k,i}) + \sigma_{{\rm R}_k}^2 }\right).
\end{align}
Note that in \eqref{AchievableErgodicRate}, $| \mathbb{E}\{ \bm{h}_{k,i}^H \bm{b}_{k,i} \} |^2$ and $\mathrm{Var}(\bm{h}_{k,i}^H \bm{b}_{k,i})$ are independent of the time block index $i$ due to the fact that the channel coefficients are i.i.d. over different time blocks.
Based on \eqref{AchievableErgodicRate}, we consider to maximize the sum of the ergodic rates by optimizing the power allocation and the pilot length, i.e,
\begin{align}
\label{Opt2}
\max_{\bm{p}\in \mathcal{R}_+^K, 0<  N_t < N}&\quad \bar{r} \deq \sum_{k=1}^K \bar{r}_k,
\quad
{\rm s.t.} \quad \eqref{CovertConstraint}.
\end{align}

In summary, in this section, we have formulated two optimization problems to enhance the covert communication performance.
Subject to the communication being at least $(1 - \epsilon)$-covert, we maximize the sum of the effective rates under the condition that the channels are quasi-static, and we maximize the sum of the ergodic rates under the condition that the channels are fast-varying.
In the subsequent two sections, we present our methods to solve these two problems, respectively.

\begin{remark}
In this paper, we do not impose any transmit power constraint at the transmitter.
In fact, to avoid being detected by the adversary, the transmit power is usually very small.
This means that in many situations, the power consumption does not constitute a performance bottleneck of covert communication.
Based on this viewpoint, in this paper, we assume that the transmitter has enough power budget to accomplish its transmission.
\end{remark}

\section{Covert communication over quasi-static channels}
In this section, we study the covert communication performance under the condition the channels are quasi-static.
In the following, we first analyze the covertness constraint and transform it into a mathematically tractable form. Then, we present our method to solve the effective sum-rate maximization problem.

\subsection{Covertness constraint analysis}
By assumption, the adversary independently runs its detection process in each single time block.
In the $i$-th time block, the detection problem at the adversary can be formulated as the following binary hypothesis testing problem,
\begin{align}
\label{BHTSingleTimeBlock}
\bm{W}(i)\triangleq (\bm{w}_{1}(i),\bm{w}_{2}(i),\cdots,\bm{w}_{K}(i)) \sim
\left\{
\begin{aligned}
&f_1^{(N_d)},&&\mathcal{H}_1, \\
&f_0^{(N_d)},&&\mathcal{H}_0,
\end{aligned}
\right.
\end{align}
where $\bm{w}_{k}(i)$, $1\leq k\leq K$, are defined in \eqref{AdversarySignalModel}, $f_1^{(N_d)}$ and $f_0^{(N_d)}$ are the PDFs of $\bm{W}(i)$  under the condition that transmission occurs and does not occur in the $i$-th time block, respectively.
For notation simplicity, we omit the time block index $i$ in the subsequent part of this section.

Define $U_k \triangleq P_{k}S_{{\rm A}, {\rm T}} |\bm{g}_{k}^H \bm{b}_{k}|^2 + Q_k S_{{\rm A}, {\rm J}} | \varphi_{k} |^2  + \sigma_{{\rm A}_k}^2$  and
$V_k \triangleq Q_k S_{{\rm A}, {\rm J}}  | \varphi_{k} |^2  + \sigma_{{\rm A}_k}^2$, and then
$f_1^{(N_d)}$ and $f_0^{(N_d)}$ can be written as
\begin{align}
\label{ProbabilityDensityFunction}
f_1^{(N_d)}(\bm{W})= \prod_{k=1}^K f_{1,k}^{(N_d)}(\bm{w}_k), ~f_{0}^{(N_d)}(\bm{W})= \prod_{k=1}^K f_{0,k}^{(N_d)}(\bm{w}_k),
\end{align}
where
$f_{1,k}^{(N_d)}(\cdot) \deq \mathbb{E}_{U_k}\left\{f^{(N_d)}(\cdot|U_k)\right\}$,
$f_{0,k}^{(N_d)}(\cdot) \deq \mathbb{E}_{V_k}\left\{f^{(N_d)}(\cdot|V_k)\right\}$,
and for $x > 0$, $f^{(N_d)}(\bm{w}_k | x )  \triangleq \frac{1}{\pi^{N_d} x^{N_d}}\e^{-\frac{\bm{w}_k^H\bm{w}_k}{x}}$ is the PDF of a circular symmetrical complex Gaussian random vector with covariance matrix $x\bm{I}_{N_d}$. The PDFs of $U_k$ and $V_k$ are presented in the following lemma.
\begin{lemma}
\label{Pro:PDFUV}
The PDFs of $U_k$ and $V_k$ are given by
\begin{subequations}
\label{PDFFUandFV}
\begin{align}
\label{PDFFU}
f_{U_k}(x) &= \left\{\begin{aligned}
&\frac{\e^{-\frac{1}{\hat{q}_{k}}(x - \sigma_{{\rm A}_k}^2)} -  \e^{-\frac{1}{\hat{p}_{k}}(x - \sigma_{{\rm A}_k}^2)}}{\hat{q}_{k} - \hat{p}_{k}},~\textrm{ if }\hat{p}_{k} \neq \hat{q}_{k},\\
&\frac{x-\sigma_{{\rm A}_k}^2}{\hat{q}_{k}^2} \e^{-\frac{1}{\hat{q}_{k}}(x - \sigma_{{\rm A}_k}^2)},~\textrm{ if } \hat{p}_{k} = \hat{q}_{k},
\end{aligned}\right.\\
\label{PDFFV}
f_{V_k}(x) &= \frac{1}{\hat{q}_{k}}\e^{-\frac{1}{\hat{q}_{k}}(x - \sigma_{{\rm A}_k}^2)},
\end{align}
\end{subequations}
where $x\geq \sigma_{{\rm A}_k}^2$, $\hat{q}_{k}\triangleq Q_k S_{\rm A,J}$ and $\hat{p}_k \triangleq P_k S_{\rm A,T}$.
\end{lemma}
\begin{IEEEproof}
First of all, we have $| \varphi_{k} |^2\sim \mathcal{E}(1)$ by assumption. Besides, due to the fact that $\bm{g}_{k}\sim\mathcal{CN}(\bm{0},\bm{I}_M)$ and that $\bm{b}_{k}$ is independent of $\bm{g}_{k}$ with $|| \bm{b}_{k}||^2 = 1 $, it is straight that $\bm{g}_{k}^H \bm{b}_{k} \sim \mathcal{CN}(0,1)$ and thus $ |\bm{g}_{k}^H \bm{b}_{k}|^2 \sim \mathcal{E}(1)$.
\end{IEEEproof}

Note that in covert communication, the signal power is usually very small compared to the jamming-plus-noise power in order to guarantee the covertness of the transmission.
Therefore, in the following, we assume that $\hat{p}_k < \hat{q}_{k} $ holds.

By \cite[Theorem 13.1.1]{E.Lehmann}, the optimal detector of the adversary that minimize the sum pf the false alarm and missed detection probabilities satisfies
$\min_d~\mathbb{P}_{\rm FA}(d) + \mathrm{P}_{\rm MD}(d) = 1 - \mathbb{V}( f_1^{(N_d)} , f_0^{(N_d)} )$.
Therefore, the covertness constraint \eqref{CovertConstraint} can be re-formulated as $\mathbb{V}( f_1^{(N_d)} , f_0^{(N_d)} ) \leq \epsilon$.
The following proposition presents an expression for $\mathbb{V}( f_1^{(N_d)} , f_0^{(N_d)} )$ given that $N_d$ is sufficiently large.
\begin{proposition}
\label{Pro:TVnri}
Define $f_{U}(\bm{u}) \triangleq \prod_{k=1}^K f_{U_k}(u_k) $ and $f_{V}(\bm{v}) \triangleq \prod_{k=1}^K f_{V_k}(v_k) $, where $f_{U_k}$ and $f_{V_k} $, for $1\leq k\leq K$, are presented in \eqref{PDFFU} and \eqref{PDFFV}, respectively.
As $N_d \ri $, the total variation distance between $f_1^{(N_d)}$ and $f_0^{(N_d)}$ satisfies
$\lim_{N_d \rightarrow \infty }\mathbb{V}\left( f_1^{(N_d)} , f_0^{(N_d)} \right)
= \mathbb{V}\left( f_U , f_{V} \right)$.
\end{proposition}
\begin{IEEEproof}
Please refer to Appendix \ref{APP:Pro1}.
\end{IEEEproof}

For the special case of $K = 1$, $\mathbb{V}\left( f_U , f_{V} \right)$ can be written in a closed form,
\begin{align}
& \mathbb{V}\left( f_U , f_{V} \right) =
\frac{1}{2}\int_{\sigma_{{\rm A}_1}^2}^{\infty} |f_{U_1}(x) -  f_{V_1}(x)| \rmd x \nonumber \\
=&
\frac{1}{2} \int_{\mathcal{X}_1} f_{U_1}(x) -  f_{V_1}(x) \rmd x + \frac{1}{2}\int_{\mathcal{X}_2} f_{V_1}(x) -  f_{U_1}(x) \rmd x\nonumber\\
=&
\frac{1}{2} \int_{\mathcal{X}_1} f_{U_1}(x) -  f_{V_1}(x) \rmd x + \frac{1}{2}\int_{\mathcal{X}_2} f_{V_1}(x) -  f_{U_1}(x) \rmd x\nonumber \\
&+ \frac{1}{2} \int_{\mathcal{X}_1} f_{U_1}(x) -  f_{U_1}(x) \rmd x + \frac{1}{2} \int_{\mathcal{X}_1} f_{V_1}(x) -  f_{V_1}(x) \rmd x \nonumber\\
\overset{(a)}{=}& \int_{\mathcal{X}_1} f_{U_1}(x) -  f_{V_1}(x) \rmd x  \label{K1TVaMiddleStep}\\
\overset{(b)}{=}& \int_{\frac{\hat{q}_1 \hat{p}_1}{\hat{q}_{1} - \hat{p}_1} \ln\frac{\hat{q}_{1}}{\hat{p}_1}}^{\infty} \frac{\hat{p}_1\e^{-\frac{1}{\hat{q}_{1}}x}}{(\hat{q}_{1} - \hat{p}_1)\hat{q}_{1}}-  \frac{ \e^{-\frac{1}{\hat{p}_{1}}x}}{\hat{q}_{1} - \hat{p}_1} \rmd x
=  \chi_1^{\frac{1}{1 - \chi_1} }. \label{K1TV}
\end{align}
where
$\mathcal{X}_1 \deq \left\{x: f_{U_1}(x) \geq  f_{V_1}(x),x \geq \sigma_{{\rm A}_1}^2 \right\}$,
$\mathcal{X}_2 \deq \left\{x: f_{U_1}(x) <  f_{V_1}(x), x\geq \sigma_{{\rm A}_1}^2 \right\}$,
$\chi_1\deq \frac{\hat{p}_1}{\hat{q}_1}$, step $(a)$ is due to the fact that  $\int_{ \sigma_{{\rm A}_1}^2 }^{\infty} f_{U_1}(x)\rmd x = \int_{ \sigma_{{\rm A}_1}^2 }^{\infty} f_{V_1}(x)\rmd x = 1$, and step $(b)$ is obtained by substituting $f_{U_1}(x)$ and $f_{V_1}(x)$ derived in \eqref{PDFFUandFV} into \eqref{K1TVaMiddleStep} and letting $x \leftarrow x - \sigma_{{\rm A}_1}^2$.

For $K\geq 2$, it is still too complicated to simplify $\mathbb{V}\left( f_U , f_{V} \right)$, which involves a $K$-fold integral. Therefore, in the following, we derive an analytically tractable upper bound on $\mathbb{V}\left( f_U , f_{V} \right)$ for the ease of numerically optimizing the covert communication performance.
\begin{proposition}
\label{Prop:UpperBoundOnTV}
An upper bound on  $\mathbb{V}\left( f_U , f_{V} \right)$ is given by
\begin{align}
\mathbb{V}\left( f_U , f_{V} \right) \leq \sum_{k=1}^K \chi_k^{\frac{1}{1 - \chi_k} }\label{TVUpperBoundGeneralK}
\end{align}
where for $1\leq k\leq K$, $\chi_k\deq \frac{\hat{p}_k}{\hat{q}_k} = \frac{P_k S_{\rm A,T}}{Q_k S_{\rm A,J}}$.
\end{proposition}
\begin{IEEEproof}
By the definition of $\mathbb{V}\left( f_U , f_{V} \right)$, we have
\begin{align}
&\mathbb{V}\left( f_U , f_{V} \right) = \mathbb{V}\left( \prod_{k=1}^K f_{U_k} , \prod_{k=1}^K f_{V_k} \right) \nonumber \\
=& \frac{1}{2}\int_{\Upsilon} \left|  \prod_{k=1}^K f_{U_k}(x_k)  - \prod_{k=1}^K f_{V_k}(x_k) \right|\rmd \bm{x} \nonumber \\
=& \frac{1}{2}\int_\Upsilon \Bigg| \left(f_{U_k}(x_1)-f_{V_k}(x_1)\right) \left(  \prod_{k=2}^K f_{U_k}(x_k) \right) \nonumber \\
& + f_{V_k}(x_1)\left(\prod_{k=2}^K f_{U_k}(x_k) - \prod_{k=2}^K f_{V_k}(x_k)\right) \Bigg|\rmd \bm{x} \nonumber \\
\overset{(a)}{\leq} & \frac{1}{2}\int_\Upsilon \left| \left(f_{U_k}(x_1)-f_{V_k}(x_1)\right) \prod_{k=2}^K f_{U_k}(x_k)\right|\rmd \bm{x} \nonumber \\
&+\frac{1}{2}\int_\Upsilon \left| f_{V_k}(x_1) \left(\prod_{k=2}^K f_{U_k}(x_k) - \prod_{k=2}^K f_{V_k}(x_k)\right) \right|\rmd \bm{x} \nonumber \\
\overset{(b)}{=}
& \mathbb{V}\left( f_{U_1} , f_{V_1} \right) + \mathbb{V}\left( \prod_{k=2}^K f_{U_k} , \prod_{k=2}^K f_{V_k} \right)\nonumber \\
\overset{(c)}{=} & \chi_1^{\frac{1}{1 - \chi_1} } + \mathbb{V}\left( \prod_{k=2}^K f_{U_k} , \prod_{k=2}^K f_{V_k} \right) \nonumber \\
\overset{(d)}{\leq}  & \chi_1^{\frac{1}{1 - \chi_1} } + \chi_2^{\frac{1}{1 - \chi_2} } + \mathbb{V}\left( \prod_{k=3}^K f_{U_k} , \prod_{k=3}^K f_{V_k} \right) \nonumber \\
\leq &  \cdots \leq \sum_{k=1}^K \chi_k^{\frac{1}{1 - \chi_k}} \label{TVUpperBoundStaticChannel}
\end{align}
where $\bm{x}=[x_1,x_2,\cdots,x_K]^T$,
$\Upsilon \deq \{ \bm{x}: x_k\geq \sigma_{{\rm A}_k}^2 \text{ for } 1\leq k\leq K\}$,
$\Upsilon^{\prime} \deq \{ [x_2,x_3,\cdots,x_K]^T: x_k\geq \sigma_{{\rm A}_k}^2 \text{ for } 2\leq k\leq K\} $,
step $(a)$ is due to the triangle inequality, i.e., $|u+v|\leq |u|+|v|$ for any real-valued $u$ and $v$,
step $(b)$ is obtained by calculating the integral with respect to $x_2,x_3,\cdots,x_K$ for the first term and  calculating the integral with respect to $x_1$ for the second term, and using the definition of $\mathbb{V}(\cdot,\cdot)$,
step $(c)$ follows from \eqref{K1TV},
step $(d)$ is obtained by applying the steps from $(a)$ to $(c)$ to $\mathbb{V}\left( \prod_{k=2}^K f_{U_k} , \prod_{k=2}^K f_{V_k} \right)$.
\end{IEEEproof}

Based on Proposition \ref{Prop:UpperBoundOnTV}, in order to make sure that the communication is at least $(1 - \epsilon)$-covert, we consider the following constraint,
\begin{align}
\label{EffectiveRateAppx1Cons}
\sum_{k=1}^K \eta(\chi_k) \leq  \epsilon,
\end{align}
where $\eta(x) \deq x^{\frac{1}{1 - x}}$.
The tightness of using $\sum_{k=1}^K \eta(\chi_k)$ as an upper bound on $\mathbb{V}\left( f_U , f_{V} \right)$ will be numerically evaluated in Section \ref{Sec:Num1}.

\subsection{Covert communication performance optimization}
\label{SubSec:POA}
In previous subsection, we have presented our method to reformulate the covertness constraint.
In this subsection, we maximize the effective sum-rate subject to the covertness constraint \eqref{EffectiveRateAppx1Cons}.

For the case with $K = 1$,
based on  \eqref{K1TV}, we ensure the covertness of the communication by letting $\chi_1^{\frac{1}{1 - \chi_1} }\leq \epsilon$.
As a result, the maximal feasible transmit power is given by $P_1^* = \frac{\chi^* \hat{q}_1}{S_{\rm A,T}}$, where $\chi^*$ is the unique solution to $x^{\frac{1}{1 - x} } = \epsilon$ in $(0,1)$.
Substituting $P_1^*$ into \eqref{Opt1}, we obtain the following problem,
\begin{align}
\label{SingleReceiver}
\max_{ 0\leq \gamma_1 \leq A_1 \chi^* /\ln(B_1) }
&~   R_1^{(e)}(\gamma_1)\triangleq\left(1 - B_1 \e^{ - \frac{A_1 \chi^*}{\gamma_1} } \right)\ln\left(1 + \gamma_1\right),
\end{align}
where $\gamma_1 \triangleq \e^{r_1} - 1$ is viewed as the optimization variable,
$A_1 \triangleq \frac{ S_{{\rm R}_k,{\rm T}} S_{\rm A, J} }{ S_{{\rm R}_k,\rm J} S_{\rm A, T} } ||\bm{h}_{1}||^2$ and
$B_1\triangleq \e^{\sigma_{{\rm R}_k}^2/( Q_{1}S_{{\rm R}_k,\rm J})}$.
The optimal solution to \eqref{SingleReceiver}, denoted by $\gamma_1^*$, is presented below.
\begin{proposition}
\label{FixedPowerOptimalGamma} $\gamma_1^* = \frac{1}{\kappa^*}$, where $\kappa^*$ is the unique solution in $(0,\infty)$ to  equation $\Xi(\kappa) = B_1$, where $\Xi(\kappa) \deq \e^{ A_1 \chi^* \kappa} - A_1B_1 \chi^*  \kappa ( 1 + \kappa )\ln\left(1+ \kappa^{-1}\right)$.
\end{proposition}
\begin{IEEEproof}
Let $\kappa = \frac{1}{\gamma_1}$ and denote $\hat{R}(\kappa) = R_1^{(e)}(\kappa^{-1})$. Now, we maximize $\hat{R}(\kappa)$ with respect to $\kappa$.
First of all, we have $\hat{R}^{\prime}(\kappa) = \frac{\e^{-A_1 P_1^* \kappa}\hat{\Xi}(\kappa)}{\kappa(1+\kappa)}$ where $\hat{\Xi}(\kappa) \deq B_1 - \Xi(\kappa)$.
It can be seen that the sign of $\hat{R}(\kappa)$ is determined by that of $\hat{\Xi}(\kappa)$.
By checking the derivative of $\hat{\Xi}(\kappa)$, it can be easily shown that $\hat{\Xi}(\kappa)$ first increases and then decreases with $\kappa$. Since $\hat{\Xi}(0) = B_1 - 1 > 0$ and $\lim_{\kappa\ri } \hat{\Xi}(\kappa) = -\infty$, we conclude that $\kappa^*$, which is the root of $\hat{\Xi}(\kappa) = 0$, maximizes the value of $\hat{R}(\kappa)$. It is worth noting that $\kappa^* \geq \frac{\ln(B_1)}{A_1P_1^*}$ always holds. Otherwise, $\hat{R}(\kappa^*)\leq 0$ by definition. Therefore, $\gamma_1^* = \frac{1}{\kappa^*}$ is the unique solution to \eqref{SingleReceiver}.
\end{IEEEproof}

Now, we discuss the case where $K\geq 2$. By using $\sum_{k=1}^K \eta(\chi_k)$ as an upper approximation of $\mathbb{V}\left( f_U , f_{V} \right)$, we obtain the following optimization problem,
\begin{subequations}
\label{EffectiveRateAppx1}
\begin{align}
\max_{\bm{\chi} ,\bm{\gamma}\in \mathcal{R}_+^K}&\quad  R^{(e)}(\bm{\chi},\bm{\gamma}), \\
{\rm s.t.}~~&\quad \sum_{k=1}^K \eta(\chi_k) \leq \epsilon,
\label{EffectiveRateAppx1Final}
\end{align}
\end{subequations}
where
$R^{(e)}(\bm{\chi},\bm{\gamma}) \deq \sum_{k=1}^K \mathbb{I}\{B_k \e^{ - \frac{A_k \chi_k}{\gamma_k}} \leq 1\}(1 - B_k \e^{ - \frac{A_k \chi_k}{\gamma_k}} )\ln(1 + \gamma_k)$,
$\bm{\chi} = [\chi_1,\chi_2,\cdots,\chi_K]^T$ with $\chi_k \triangleq \frac{P_kS_{\rm A,T}}{Q_kS_{\rm A,J}}$,
$\bm{\gamma} = [\gamma_1,\gamma_2,\cdots,\gamma_K]^T$ with $\gamma_k \triangleq \e^{r_k} - 1$,
$A_k \triangleq \frac{S_{{\rm R}_k,{\rm T}}S_{\rm A,J}}{S_{{\rm R}_k,{\rm J}}S_{\rm A,T}} ||\bm{h}_{k}||^2 $,
and $B_k \triangleq \e^{\sigma_{{\rm R}_k}^2/(Q_k S_{{\rm R}_k,{\rm J}})}$.
Note that in \eqref{EffectiveRateAppx1}, we view $(\bm{\chi},\bm{\gamma})$ as the optimization variables instead of $(\bm{p},\bm{r})$ for  convenience.

Now, we present our method to solve \eqref{EffectiveRateAppx1}.
Though \eqref{EffectiveRateAppx1} is a non-convex optimization problem, its global optimal solution can still be obtained.
First of all, we observe that for a fixed $\bm{\chi}$, \eqref{EffectiveRateAppx1} becomes $K$ parallel sub-problems. Specifically,  the $k$-th sub-problem, with a single optimization variable $\gamma_k$, is in the same form as \eqref{SingleReceiver}.
Therefore, for a fixed $\bm{\chi}$, we can obtain the optimal $\bm{\gamma}$, denoted by $\bm{\gamma}^*(\bm{\chi})$, by using Proposition \ref{FixedPowerOptimalGamma}.
In this way, \eqref{EffectiveRateAppx1} can be reformulated as
\begin{align}
\max_{\bm{\chi}\in \mathcal{R}_+^K}~\tilde{R}^{(e)}(\bm{\chi}),
\quad {\rm s.t.}~\eqref{EffectiveRateAppx1Final}
\label{EffectiveRateAppx2}
\end{align}
where $\tilde{R}^{(e)}(\bm{\chi}) \deq R^{(e)}(\bm{p},\bm{\gamma}^*(\bm{\chi}) ) $.
For optimization problem \eqref{EffectiveRateAppx2}, we observe that:
\begin{enumerate}
\item $\tilde{R}^{(e)}(\bm{\chi})$ is a monotonically increasing function (see Definition \ref{D:Increasing} in Appendix \ref{APP:POA}); specifically, for any two points $\bm{\chi}_1$ and $\bm{\chi}_2$, if $\bm{\chi}_1\geq\bm{\chi}_2\geq0$, then $\tilde{R}^{(e)}(\bm{\chi}_1) \geq \tilde{R}^{(e)}(\bm{\chi}_2)$;
\item the feasible set of $\bm{\chi}$, denoted by $\mathcal{X}$,  is contained in a box (see Definition \ref{D:Box} in Appendix \ref{APP:POA}), i.e., for $\forall \bm{\chi}\in \mathcal{X}$, it satisfies that $\bm{0}\leq \bm{\chi} \leq \bar{\bm{\chi}} =\chi^* \bm{1}_K$, which follows from \eqref{EffectiveRateAppx1Final};
\item the feasible set  $\mathcal{X}$ is a normal set (see Definition \ref{D:NormalSet} in Appendix \ref{APP:POA}); specifically, for any two points $\bm{\chi}_1\geq \bm{0}$ and $\bm{\chi}_2\geq\bm{0}$, if $\bm{\chi}_1\leq \bm{\chi}_2$ and $\bm{\chi}_2\in\mathcal{X}$, then $\bm{\chi}_1\in\mathcal{X}$;
\end{enumerate}
Based on these observations, optimization problem \eqref{EffectiveRateAppx1} is a \emph{monotonic optimization problem}, and its globally optimal solution can be obtained by using the POA method \cite{Y.J.Zhang2012,H.Tuy2005}.
We present a brief introduction on the POA method in Appendix \ref{APP:POA}.

\subsection{An SCA based method}
\label{SubSec:SCA}
In this subsection, we present an alternative method to search for a suboptimal solution to \eqref{EffectiveRateAppx1}.
Define $\bm{t} = [t_1,t_2,\cdots,t_K]^T$ with $t_k \deq \frac{\chi_k}{\gamma_k}$ for $1\leq k\leq K$. We view $\bm{t}$ and $\bm{\gamma}$ as the optimization variables and recast \eqref{EffectiveRateAppx1} as the following problem,
\begin{subequations}\label{SCAReformulate1}
\begin{align}
\max_{\bm{t},\bm{\gamma}\in \mathcal{R}_+^K}
&~~\sum_{k=1}^K \left(1 - B_k \e^{ - A_k t_k } \right)\ln\left(1 + \gamma_k\right)\\
{\rm s.t.}~~&~~\sum_{k=1}^K \eta(t_k \gamma_k) \leq \epsilon, & \label{SCAReformulate1C1}\\
&~~B_k \e^{ - A_k t_k } \leq 1,~~1\leq k\leq K, & \label{SCAReformulate1C2}
\end{align}
\end{subequations}
The main difficulty on solving \eqref{SCAReformulate1} lies in the fact that \eqref{SCAReformulate1C1} is not a convex constraint and that the objective of \eqref{SCAReformulate1} is non-concave.
Note that $\eta(x)$ satisfies that
$\eta(0) = 0$,
$\lim_{x\rightarrow 0}\eta^{\prime}(x) = 1$.
By checking the high order derivatives of $\eta(x)$, if can be proved that $\eta^{\prime\prime}(x)\leq 0$ for $\forall x\in (0,1)$.
Therefore, $\eta(x)$ is a monotonically increasing concave function, and satisfies
\begin{align}
\eta(x) \leq x,\quad \forall x>0 \label{NewCovertConstraintBound}
\end{align}
Based on \eqref{NewCovertConstraintBound}, we can replace the inequality constraint \eqref{SCAReformulate1C1} with $\bm{t}^T\bm{\gamma}\leq \epsilon$ to obtain a lower bound on optimal covert communication performance.
We further simplify \eqref{SCAReformulate1},
by  introducing two vector-valued slack variables $\bm{\alpha}=[\alpha_1,\alpha_2,\cdots,\alpha_K]^T\in \mathcal{R}_+^K$ and $\bm{\beta}=[\beta_1,\beta_2,\cdots,\beta_K]^T\in \mathcal{R}_+^K$,   and obtain the following problem,
\begin{subequations}\label{SCAStep1}
\begin{align}
\min_{\bm{t},\bm{\gamma},\bm{\alpha},\bm{\beta}\in \mathcal{R}_{+}^K}&~~R(\bm{\alpha},\bm{\beta})\\
{\rm s.t.}~~~~&~~\bm{t}^T\bm{\gamma}\leq \epsilon\, \label{SCAStep1C1}\\
&~~1 - B_k \e^{ - A_k t_k } \geq  \alpha_k,~1\leq k\leq K, \label{SCAStep1C2}\\
&~~\ln\left(1 + \gamma_k\right) \geq \beta_k,~1\leq k\leq K, \label{SCAStep1C3}
\end{align}
\end{subequations}
where $R(\bm{\alpha},\bm{\beta}) \triangleq \frac{||\bm{\alpha} - \bm{\beta}||^2-||\bm{\alpha} + \bm{\beta}||^2}{4} = -\bm{\alpha}^T\bm{\beta}$. Note that in \eqref{SCAStep1}, constraint \eqref{SCAReformulate1C2} is neglected because \eqref{SCAReformulate1C2} is guaranteed if \eqref{SCAStep1C2} is satisfied and $\bm{\alpha}\in \mathcal{R}_{+}^K$.

Problem \eqref{SCAStep1} is still a non-convex problem due to the non-convex objective function $R_s(\bm{\alpha},\bm{\beta})$ and the non-convex constraint \eqref{SCAStep1C1}. However, \eqref{SCAStep1}, in its current form, can be efficiently handled by the SCA method. In brief, the SCA method handles a non-convex optimization problem by transforming it into a series of parameterized convex problem. By iteratively solving the obtained convex problems with the parameters at each iteration being the optimal solution obtained in previous iteration, the SCA method generates a series of solutions which converges to a Karush-Kuhn-Tucker (KKT) point of the original non-convex problem. A brief introduction of the SCA method is presented in Appendix \ref{APP:SCA}. In our case, in the $(j+1)$-th iteration of the SCA method, we need to solve the following convex problem,
\begin{subequations}\label{SCAStep2}
\begin{align}
\min_{\bm{t},\bm{\gamma},\bm{\alpha},\bm{\beta}\in \mathcal{R}_{+}}&
||\bm{\alpha} - \bm{\beta}||^2 - 2\left(\bm{\rho}^{(j)}\right)^T(\bm{\alpha} + \bm{\beta})\\
{\rm s.t.}~~~~&\frac{1}{2}\bm{t}^T\bm{\Lambda}_1^{(j)}\bm{t} + \frac{1}{2}\bm{\gamma}^T\bm{\Lambda}_2^{(j)}\bm{\gamma} \leq \epsilon, \label{SCAStep2C1}\\
&1 - B_k \e^{ - A_k t_k } \geq  \alpha_k,~1\leq k\leq K , \label{SCAStep2C2}\\
&\ln\left(1 + \gamma_k\right) \geq \beta_k,~1\leq k\leq K , \label{SCAStep2C3}
\end{align}
\end{subequations}
where $\bm{\rho}^{(j)} = \bm{\alpha}^{(j)} + \bm{\beta}^{(j)}$,
$\bm{\Lambda}_1^{(j)} = \left(\bm{\Gamma}^{(j)}\right)^{-1}\bm{T}^{(j)}$,
$\bm{\Lambda}_2^{(j)} = \left(\bm{T}^{(j)}\right)^{-1}\bm{\Gamma}^{(j)}$,
$\bm{\Gamma}^{(j)} = {\rm diag}(\bm{\gamma}^{(j)})$,
$\bm{T}^{(j)} = {\rm diag}(\bm{t}^{(j)})$,
and $\{\bm{\alpha}^{(j)}, \bm{\beta}^{(j)} ,  \bm{t}^{(j)} , \bm{\gamma}^{(j)}\}$ consists of the optimal solution obtained in the $j$-th iteration. The derivation of \eqref{SCAStep2} is explained in Appendix \ref{APP:SCA}. To start the SCA iteration, $\{\bm{\alpha}^{(0)}, \bm{\beta}^{(0)} ,  \bm{t}^{(0)} , \bm{\gamma}^{(0)}\}$ can be initialized to be any feasible solution to \eqref{SCAStep1}.
Note that as \eqref{SCAStep2} is a convex problem, it can be efficiently solved by using software such as CVX \cite{CVX}.

\begin{remark}
\label{ComplexityPOASCA}
In this section, we have proposed the POA and the SCA methods to maximize the effective sum-rate subject to a covertness constraint.
Though the POA method provisdes us the globally optimal solution, it generally exhibits a high computational complexity.
Specifically, after a vertex is searched, $K$ new vertexes will be added into the vertex set (see Appendix \ref{APP:POA} for more details), meaning that the worst-case computational complexity of the POA method scales with $K$ exponentially.
As for the SCA method, it solves a sequence of convex optimization problems, i.e., \eqref{SCAStep2}, by using standard convex optimization algorithms, for example, the interior-point method combined with Newton’s method \cite{S.Boyd}, and the computational complexity grows with $K$ in a polynomial manner.
Based on these observations, we conclude that the POA method better suits to the case where the number of receivers, $K$, is small, whereas if $K$ is large, the SCA is computationally more efficient.
\end{remark}

\section{Covert communication over Fast-Varying Channels}
In this section, we consider the situation where the wireless channels are fast-varying. In the following, we first analyze the covertness constraint, based on which we then discuss the ergodic sum-rate maximization problem.

\subsection{Covertness constraint analysis}
Without loss of generality, assume that the transmission starts in the $1$-st and ends at the $L$-th time blocks (if it occurs). Accordingly, the detection problem  of the adversary  can be formulated as the following hypothesis testing problem,
\begin{align}
\label{BHTMultiTimeBlock}
\tilde{\bm{W}}  = (\bm{W}(1),\bm{W}(2),\cdots,\bm{W}(L)) \sim
\left\{
\begin{aligned}
&\tilde{f}_1^{(N_d)} ,&&\mathcal{H}_1, \\
&\tilde{f}_0^{(N_d)} ,&&\mathcal{H}_0,
\end{aligned}
\right.
\end{align}
where $\bm{W}(i)$, $1\leq i\leq L$, are defined in \eqref{BHTSingleTimeBlock},
$\tilde{f}_s^{(N_d)}(\tilde{\bm{W}}) \deq \prod_{i=1}^L f_s^{(N_d)}(\bm{W}(i))$ for $s \in \{0,1\}$ are the PDFs of
$\tilde{\bm{W}}$ under $\mathcal{H}_1$ and $\mathcal{H}_0$, respectively, with $f_1^{(N_d)}$ and $f_0^{(N_d)}$  defined in \eqref{ProbabilityDensityFunction}.
Based on \eqref{BHTMultiTimeBlock}, by using  \cite[Theorem 13.1.1]{E.Lehmann}, we obtain that
$\min_d~\mathbb{P}_{\rm FA}(d) + \mathrm{P}_{\rm MD}(d) = 1 - \mathbb{V}( \tilde{f}_1^{(N_d)}, \tilde{f}_0^{(N_d)} )$.
Note that \eqref{BHTMultiTimeBlock} differs from \eqref{BHTSingleTimeBlock} in the sense that  in \eqref{BHTMultiTimeBlock}, the signal matrices obtained in different time blocks are collected together to make a decision, while in \eqref{BHTSingleTimeBlock}, the decision is made based solely on the signal matrix obtained in a single time block.

In principle, we can use the method in Proposition \ref{Pro:TVnri} and \ref{Prop:UpperBoundOnTV} to obtain an upper bound on $\mathbb{V}( \tilde{f}_1^{(N_d)}, \tilde{f}_0^{(N_d)} )$. However, given that the channel coherent time is short, such an upper bound, which requires $N_d\rightarrow \infty$, becomes inaccurate.
In the following, we use Pinsker's inequality, see e.g., \cite[Lemma 11.6.1]{Information} to derive a mathematically tractable bound on $\mathbb{V}( \tilde{f}_1^{(N_d)}, \tilde{f}_0^{(N_d)} )$. We have
\begin{align}
\label{pinskinequality}
\mathbb{V}( \tilde{f}_1^{(N_d)}, \tilde{f}_0^{(N_d)} ) &\leq \sqrt{\frac{1}{2} \mathbb{D} (\tilde{f}_0^{(N_d)}, \tilde{f}_1^{(N_d)})  }\nonumber \\
&\overset{(a)}{=}\sqrt{\frac{ L }{2}\sum_{k=1}^K \mathbb{D} \left(f_{0,k}^{(N_d)}, f_{1,k}^{(N_d)}\right) },
\end{align}
where step $(a)$ is because $\bm{W}(i)$ for $1\leq i\leq L$ are i.i.d. random matrices.
Based on \eqref{pinskinequality}, we use the following constraint to ensure the covertness of the transmission,
\begin{align}
\label{KLBoundShort}
\sum_{k=1}^K \mathbb{D} \left(f_{0,k}^{(N_d)}, f_{1,k}^{(N_d)}\right) \leq \frac{2}{L}\epsilon^2.
\end{align}
The mathematical expression of $\mathbb{D} \left(f_{0,k}^{(N_d)}, f_{1,k}^{(N_d)}\right)$ is presented in the following proposition
\begin{proposition}
\label{Pro:KL}
For $n\geq 1$, $\mathbb{D} (f_{0,k}^{(n)}, f_{1,k}^{(n)})  = -\int_{0}^{\infty} \frac{z^{n-1}\Phi(q_k , z) }{(n-1)!}  \ln\  \Psi(p_k,q_k,z) \rmd z$,
where $\Psi(p_k,q_k,z) \triangleq 1 + \frac{p_k}{q_k - p_k}\left(1 - \frac{\Phi(p_k,z)}{\Phi(q_k ,z)}\right)$,
$\Phi(x,z) \triangleq \int_{0}^{\infty}\e^{-v} \frac{1}{ (1+x v)^n}\e^{-\frac{z}{1+x v}} \rmd v$,
$p_k \triangleq \frac{\hat{p}_k}{\sigma_{{\rm A}_k}^2}$, and $q_k \triangleq \frac{\hat{q}_k}{\sigma_{{\rm A}_k}^2}$.
\end{proposition}
\begin{IEEEproof}
By definition, $\mathbb{D} (f_{0,k}^{(n)}, f_{1,k}^{(n)})$ can be simplified as \eqref{OriginalKLDiv}
in the top of next page.
\begin{figure*}[bht!]
\begin{align}
& \mathbb{D} (f_{0,k}^{(n)}, f_{1,k}^{(n)}) =  \int_{\mathcal{C}^{n}}
\mathbb{E}_{V_k}\left\{ \frac{V_k^{-n}\e^{-\bm{w}^H\bm{w}/V_k}}{\pi^{n} }\right\}
\ln
\frac
{ \mathbb{E}_{V_k} \{V_k^{-n}\e^{-\bm{w}^H\bm{w}/V_k}  \} }
{ \mathbb{E}_{U_k} \{U_k^{-n}\e^{-\bm{w}^H\bm{w}/U_k}  \} } \rmd \bm{w} \nonumber \\
\overset{(a)}{=}& \int_{\mathcal{R}_+^{n}}\left(\prod_{i=1}^n x_i \right)\int_{(-\pi,\pi]^n}
\mathbb{E}_{V_k}\left\{\frac{V_k^{-n} \e^{-\bm{x}^T\bm{x}/V_k}}{\pi^{n} }\right\}
\ln
\frac
{ \mathbb{E}_{V_k} \{V_k^{-n}\e^{-\bm{x}^T\bm{x}/V_k}  \} }
{ \mathbb{E}_{U_k} \{U_k^{-n}\e^{-\bm{x}^T\bm{x}/U_k}  \} } \rmd \bm{\theta}\rmd \bm{x} \nonumber \\
\overset{(b)}{=}& \int_{\mathcal{R}_+^{n}}
\mathbb{E}_{V_k}\left\{V_k^{-n}\e^{-\sum_{i=1}^n y_i/V_k}\right\}
\ln
\frac
{ \mathbb{E}_{V_k} \{V_k^{-n}\e^{-\sum_{i=1}^n y_i/V_k}  \} }
{ \mathbb{E}_{U_k} \{U_k^{-n}\e^{-\sum_{i=1}^n y_i/U_k}  \} } \rmd \bm{y} \nonumber \\
\overset{(c)}{=}& \int_{\mathcal{R}_+} \frac{z^{n-1}}{(n-1)!}
\mathbb{E}_{V_k}\left\{V_k^{-n}\e^{-z/V_k}\right\}
\ln
\frac
{ \mathbb{E}_{V_k} \{V_k^{-n}\e^{-z/V_k}  \} }
{ \mathbb{E}_{U_k} \{U_k^{-n}\e^{-z/U_k}  \} } \rmd z \label{OriginalKLDiv}
\end{align}
\noindent\rule[0.25\baselineskip]{\textwidth}{1pt}
\end{figure*}
where in step $(a)$, we express the complex-valued integral variables $w_1,w_2,\cdots,w_n$ in polar coordinates, i.e.,
$w_i = x_i \e^{\mathrm{j} \theta_i}$ with $x_i\in \mathcal{R}_+$ and $\theta_i \in (-\pi,\pi]$,
step $(b)$ is obtained by first calculating the integral w.r.t.  $\bm{\theta}$ and
then changing the integral variables by letting $y_i = x_i^2$ for $1\leq i \leq n$;
and finally, in step $(c)$, we let $z = \sum_{j=1}^n y_j$ and $z_i = \sum_{j=1}^i y_j$ for $1\leq i\leq n-1$, and calculate the integral w.r.t. $(z_1,z_2,\cdots,z_{n-1})$.
We further have that
\begin{align}
&\mathbb{E}_{V_k} \left\{\frac{\e^{-\frac{z}{V_k}}}{V_k^{n}}\right\}
= \int_{\sigma_{{\rm A}_k}^2}^{\infty} \frac{\e^{-\frac{v-\sigma_{{\rm A}_k}^2}{\hat{q}_k} }}{\hat{q}_k}\frac{ \e^{-\frac{z}{v}} }{v^n} \rmd v \nonumber \\
=& \frac{1}{\sigma_{{\rm A}_k}^{2n}}\int_{0}^{\infty}  \frac{ \e^{- v}\e^{-\frac{z/\sigma_{{\rm A}_k}^2}{1 + q_k v}}}{( 1 + q_{k} v)^n}
 \rmd v  = \frac{1}{\sigma_{{\rm A}_k}^{2n}} \Phi\left(q_k,\frac{z}{\sigma_{{\rm A}_k}^2}\right).
\label{PHIV}
\end{align}
Similarly, we obtain
\begin{align}
\mathbb{E}_{U_k} \left\{\frac{\e^{-\frac{z}{U_k}}}{U_k^{n}}\right\}
=&  \frac{1}{\sigma_{{\rm A}_k}^{2n}}\Bigg( \frac{q_k}{q_k - p_k}\Phi\left(q_k,\frac{z}{\sigma_{a,k}^2}\right) \nonumber \\
&\quad - \frac{p_k}{q_k - p_k}\Phi\left(p_k,\frac{z}{\sigma_{{\rm A}_k}^2}\right)\Bigg).
\label{PHIU}
\end{align}
We complete the proof by inserting \eqref{PHIV} and \eqref{PHIU} into \eqref{OriginalKLDiv} and letting $z\leftarrow \frac{z}{\sigma_{{\rm A}_k}^2}$,
\end{IEEEproof}

Based on Proposition \ref{Pro:KL}, for a given power allocation vector $\bm{p} = [P_1,P_2,\cdots,P_K]^T$, we can check its feasibility subject to constraint \eqref{KLBoundShort}.
However,  \eqref{KLBoundShort} is too complicated to facilitate to solve optimization problem \eqref{Opt2}.
In view of the fact that $p_k$ is generally small to ensure the covertness,
the following proposition provides a way to approximate $\mathbb{D} (f_{0,k}^{(n)}, f_{1,k}^{(n)})$.
\begin{proposition}
\label{LocalQuadratic}
$\lim_{p_k \rightarrow 0}\frac{ \mathbb{D} (f_{0,k}^{(n)}, f_{1,k}^{(n)}) }{ p_k^2 } = \frac{\zeta(q_k,n)}{2 q_k^2}$, where $\zeta(q_k,n)\deq   -1 + \int_{0}^{\infty} \frac{z^{n-1}\e^{-z}}{(n-1)!}
\frac{\e^{-z}}{\Phi(q_k ,z)}\rmd z$.
\end{proposition}
\begin{IEEEproof}
First of all, it is straight that $\mathbb{D} (f_{0,k}^{(n)}, f_{1,k}^{(n)})\big|_{p_k = 0} = 0$.
Besides, it can be verified that
$\frac{\rmd \mathbb{D} (f_{0,k}^{(n)}, f_{1,k}^{(n)})}{\rmd p_k}\big|_{p_k = 0} = \int_{0}^{\infty} \frac{z^{n-1}}{(n-1)! } \frac{\e^{-z} - \Phi(q_k , z)}{q_k}\rmd z =  0$.
We further have that
$\frac{\rmd^2 \mathbb{D} (f_{0,k}^{(n)}, f_{1,k}^{(n)})}{\rmd p_k^2}\big|_{p_k = 0} = \frac{1}{q_k^2}\int_{0}^{\infty} \frac{z^{n-1}\Phi(q_k , z) }{(n-1)!} \left(1 - \frac{\e^{-z}}{\Phi(q_k ,z)}\right)^2 \rmd z = \frac{1}{q_k^2}\zeta(q_k,n)>0$, which completes the proof.
\end{IEEEproof}

Based on Proposition \ref{LocalQuadratic}, we can approximate the covertness constraint \eqref{KLBoundShort} by,
\begin{align}
\label{KLBoundShortFurther}
\sum_{k=1}^K \frac{\zeta(q_k,N - N_t)}{2 }\chi_k^2 \leq \frac{2}{L}\epsilon^2.
\end{align}
where $\chi_k$ for $1\leq k\leq K$ are defined in \eqref{EffectiveRateAppx1}.
\begin{remark}
For $q_k>0$, it can be verified through numerical calculation that $\frac{\rmd^3 \mathbb{D} (f_{0,k}^{(n)}, f_{1,k}^{(n)})}{\rmd p_k^3}\big|_{p_k = 0} < 0$, meaning that the left-hand-side (LHS) of \eqref{KLBoundShortFurther} is an upper bound on the LHS of \eqref{KLBoundShort} when $p_k$ is sufficiently small.
Note that in \eqref{KLBoundShortFurther},
with $N$ and $N_t$ fixed,
the transmit power $P_k$ is upper bounded by $O(\frac{1}{\sqrt{L}})$, which is analogous to the SRL. Note that here, $L$ stands for the number of time blocks but not the number of channel uses.
Similar result has also been presented in \cite{D.Goeckel2018SPAWC}.
\end{remark}

\subsection{Covert communication performance optimization}
\label{FCODS}
In this subsection, we present our method to maximize the sum of the ergodic rates subject to \eqref{KLBoundShortFurther}.
Based on \eqref{EstimatedChannel} and \eqref{EstimatedChannelError},  for $1\leq k\leq K$, we have
\begin{align}
| \mathbb{E}\{ \bm{h}_{k,l}^H \bm{b}_{k,l} \} |^2 &= \left| \mathbb{E}\left\{ \hat{\bm{h}}_{k,l}^H\bm{b}_{k,l} + \tilde{\bm{h}}_{k,l}^H\bm{b}_{k,l}\right\} \right|^2 \nonumber \\
& = \left| \mathbb{E}\left\{ ||\hat{\bm{h}}_{k,l}|| \right\} \right|^2 = \frac{N_t }{N_t  + \mu_k }G,\label{GG}\\
\mathrm{Var}\{ \bm{h}_{k,l}^H \bm{b}_{k,l} \} &  = \mathrm{Var}\left\{ ||\hat{\bm{h}}_{k,l}|| + \tilde{\bm{h}}_{k,l}^H\bm{b}_{k,l} \right\} \nonumber \\
&=  \frac{N_t }{N_t  + \mu_k }E + \frac{\mu_k}{N_t  + \mu_k},\label{MM}
\end{align}
where $G \triangleq \Gamma^2\left(M+\frac{1}{2}\right)/\Gamma^2\left(M\right)$, $E \triangleq  M - G$, and we have used the fact that $\hat{\bm{h}}_{k,l}$ and $\tilde{\bm{h}}_{k,l}$ are independent.
Substituting \eqref{GG} and \eqref{MM} into \eqref{AchievableErgodicRate}, we obtain,
\begin{subequations}
\label{Opt22}
\begin{align}
\max_{\bm{\chi}\in\mathcal{R}_+^K;\tau\in\mathcal{Z}_N}&\quad \bar{R}(\bm{\chi},\tau)\deq\left(1 - \tau\right)\sum_{k=1}^K  \ln\left(
1 + \overline{\rm SNR}_k \right)
 \\
{\rm s.t.}~~~~~& \quad \sum_{k=1}^K \frac{\zeta_k(\tau)}{2 }\chi_k^2 \leq \frac{2}{L}\epsilon^2. \label{Opt22Cons}
\end{align}
\end{subequations}
where $\overline{\rm SNR}_k = \frac{ \chi_k \tau G_k }{
\tau\chi_k E_k +
\chi_k \tilde{\mu}_k  +
\tau F_{k,1} +
F_{k,2}}$,
$\tau \deq \frac{N_t}{N}$,
$\mathcal{Z}_N = \left\{\frac{1}{N},\frac{2}{N},\cdots,\frac{N-1}{N}\right\} $,
$G_k\deq \frac{ Q_k S_{\rm A,J} }{S_{\rm A,T}} N G $,
$E_k \deq \frac{ Q_k S_{\rm A,J} }{S_{\rm A,T}} N E $,
$\tilde{\mu}_k\deq \frac{ Q_k S_{\rm A,J} }{S_{\rm A,T}}\mu_k$
$F_{k,1} \deq N\frac{Q_{k}S_{{\rm R}_k,{\rm J}} + \sigma_{{\rm R}_k}^2}{S_{{\rm R}_k,{\rm T}}}$,
$F_{k,2} \deq \mu_k\frac{Q_{k}S_{{\rm R}_k,{\rm J}} + \sigma_{{\rm R}_k}^2}{S_{{\rm R}_k,{\rm T}}}$,
and $\zeta_k(\tau) \deq \zeta(q_k,N(1 - \tau))$. Note that in \eqref{Opt22}, we equivalently optimize $(\bm{\chi},\tau)$ instead of $(\bm{p},N_t)$ for notation simplicity.

In the following, we present our method to solve \eqref{Opt22}.
We observe that if $\tau$ is fixed, $\bar{R}(\bm{\chi},\tau)$ is a concave function of $\bm{\chi}$, and thus \eqref{Opt22} becomes a convex optimization problem, the global optimum of which can be efficiently obtained.
In light of this, we can solve \eqref{Opt22} by first optimizing $\bm{\chi}$ with $\tau$ fixed and then calculating the optimal  $\tau$ by an exhaustively search over the finite discrete set $\mathcal{Z}_N$. For a given value of $\tau$, the optimal $\bm{\chi}$ is presented in the following proposition.
\begin{proposition}
\label{Pro:OptimalPowerFast}
Let $\tau$ be a fixed constant in $(0,1)$.
For $1\leq k\leq K$, the optimal $\chi_k$, denoted by $\chi_k^*$, is the unique solution in $(0,\infty)$ to the following equation,
\begin{align}
\chi_k \left(\tilde{G}_k \chi_k + \tilde{E}_k \chi_k + \tilde{I}_k\right)\left(\tilde{E}_k \chi_k + \tilde{F}_k \right)  = \frac{ \tilde{G}_k \tilde{F}_k }
{ \lambda^* \zeta_k(\tau) },
\label{OptimalPowerFast}
\end{align}
where $\tilde{G}_k  \deq \tau G_k$,
$\tilde{E}_k \deq \tau E_k + \tilde{\mu}_k$,
$\tilde{F}_k \deq \tau F_{k,1} + F_{k,2}$, and
$\lambda^*>0$ is the optimal Lagrange multiplier satisfying that
$\sum_{k=1}^K \frac{\zeta_k(\tau)}{2} (\chi_k^*)^2 =  \frac{2}{L}\epsilon^2$.
\end{proposition}
\begin{IEEEproof}
Let $\lambda$ be the Lagrange multiplier associated
with constraint \eqref{Opt22Cons}. The KKT conditions of \eqref{Opt22} state that
$\frac{\partial \mathcal{L}(\bm{\chi},\lambda)}{\partial \chi_k} = 0$,
where
$\mathcal{L}(\bm{\chi},\lambda) \deq \bar{R}(\bm{\chi},\tau) + \lambda\left(\sum_{k=1}^K \frac{\zeta_k(\tau)}{2}P_k^2 - \frac{2}{L}\epsilon^2\right)$ is the Lagrangian function,
which results in \eqref{OptimalPowerFast}.
Since $\bar{R}(\bm{\chi},\tau)$ and $\sum_{k=1}^K \frac{\zeta_k(\tau)}{2}\chi_k^2$ increase with $\chi_k$ for $1\leq k\leq K$, constraint \eqref{OptimalPowerFast} is active at the optimum, meaning that the optimal Lagrange multiplier $\lambda^*$ satisfies that $\sum_{k=1}^K\frac{\zeta_k(\tau)}{2}(\chi_k^*)^2 = \frac{2}{L}\epsilon^2$.
\end{IEEEproof}

Note that in Proposition \ref{Pro:OptimalPowerFast}, \eqref{OptimalPowerFast} can be solved in a closed-form due to that fact that  \eqref{OptimalPowerFast} is a cubic equation. As for $\lambda^*$, it can be efficiently searched by using the bisection method.
Based on Proposition \ref{Pro:OptimalPowerFast}, we can obtain the optimal $\bm{\chi}$ for each fixed value of $\tau$.
Then, the optimal $\tau$ can be obtained by an exhaustive search over the set $\mathcal{Z}_N$.

\subsection{An alternating optimization based method}
\label{Sec:AO}
In this subsection, we present a method to obtain a suboptimal solution of \eqref{Opt22} with a lower computational complexity than the ES based method proposed in previous subsection.
The basic idea is that we replace constraint \eqref{Opt22Cons} with the following one
\begin{align}
\label{KLBoundShortReplacement}
\sum_{k=1}^K \frac{\zeta_k(0)}{2} \chi_k^2 \leq \frac{2}{L}\epsilon^2.
\end{align}
In fact, by setting $\tau = 0$ in \eqref{Opt22Cons}, we equivalently let the adversary observe $N > N_d$ samples in each time block. Therefore, this improves the detection performance of the adversary and leads to a lower bound on the achievable communication performance.

Since \eqref{KLBoundShortReplacement} is independent of $\tau$,
we replace \eqref{Opt22Cons} with \eqref{KLBoundShortReplacement}
and solve the resulting problem by using the AO method.
Specifically, we alternatingly optimize $\bm{\chi}$ with $\tau$ fixed and optimize $\tau$ with $\bm{\chi}$ fixed until the value of $\bar{R}(\bm{\chi},\tau)$ converges.
For the sub-problem of optimizing $\bm{\chi}$, the optimal solution is presented in Proposition \ref{Pro:OptimalPowerFast}. Given that $\bm{\chi}$ is fixed, the sub-problem of optimizing $\tau$ is given by,
\begin{align}
\label{DiscreteNt}
\max_{\tau\in\mathcal{Z}_N} ~ \bar{R}_{\bm{\chi}}(\tau) \deq \left(1 - \tau \right)\sum_{k=1}^K  \ln\left(
1 + \frac{ \tau \bar{G}_k }{ \tau \bar{E}_k + \bar{F}_k }\right),
\end{align}
where $\bar{G}_k \deq \chi_k G_k$, $\bar{E}_k \deq \chi_k E_k+ F_{k,1}$, and $ \bar{F}_k \deq \chi_k \tilde{\mu}_k+ F_{k,2}$.
To solve \eqref{DiscreteNt}, we relax $\tau$ to be a real number in $(0,1)$, and the optimal real-valued $\tau$ is presented in the following proposition.
\begin{proposition}
\label{Prop:TauOptimal}
$\bar{R}_{\bm{\chi}}(\tau)$ is a concave function of $\tau$, and $\tau^* \deq \mathrm{argmax}_{\tau\in (0,1)} \bar{R}_{\bm{\chi}}(\tau)$ is the unique solution in (0,1) to the following equation
\begin{align}
\bar{R}_{\bm{\chi}}^{\prime}(\tau) &=  \sum_{k=1}^K \Bigg( \frac{\bar{G}_k\bar{F}_k(1 - \tau) }{ ( \bar{E}_k \tau + \bar{F}_k )
\left((\bar{G}_k  + \bar{E}_k )\tau + \bar{F}_k \right) } \nonumber \\
&\quad - \ln\left( 1 + \frac{ \bar{G}_k \tau }{ \bar{E}_k \tau + \bar{F}_k} \right) \Bigg)= 0. \label{OptimalTauBisection}
\end{align}
\end{proposition}
\begin{IEEEproof}
The second order derivative of $\bar{R}_{\bm{\chi}}(\tau)$ is presented in \eqref{SecondOrderDerivative} at the top of next page.
\begin{figure*}[t]
\begin{align}
\bar{R}_{\bm{\chi}}^{\prime\prime}(\tau) = -\sum_{k=1}^K
\frac{\bar{G}_k\bar{I}_k
(
2 \bar{E}_k^2 \tau + \bar{I}_k (\bar{G}_k + 2 \bar{I}_k + \bar{G}_k \tau) + 2 \bar{E}_k (\bar{I}_k + \bar{G}_k \tau + \bar{I}_k \tau)
 )
}{( \bar{E}_k \tau + \bar{I}_k )^2( (\bar{G}_k + \bar{E}_k) \tau + \bar{I}_k )^2} \label{SecondOrderDerivative}
\end{align}
\noindent\rule[0.25\baselineskip]{\textwidth}{1pt}
\end{figure*}
Given that $\tau\in (0,1)$, $\bar{R}_{\bm{\chi}}^{\prime\prime}(\tau) < 0$. Thus, $\bar{R}_{\bm{\chi}}(\tau)$ is a concave function. Besides, it can be verified that $\bar{R}_{\bm{\chi}}(0) = \bar{R}_{\bm{\chi}}(1) = 0$, $\bar{R}_{\bm{\chi}}^{\prime}(0)>0$, and $\bar{R}_{\bm{\chi}}^{\prime}(1) < 0$. Therefore, $\tau^*$ is unique, which satisfies $\bar{R}_{\bm{\chi}}^{\prime}(\tau^*) = 0$.
\end{IEEEproof}

Since $\bar{R}_{\bm{\chi}}(\tau)$ is concave, $\bar{R}_{\bm{\chi}}^{\prime}(\tau)$ decreases with $\tau$. Therefore, the root of \eqref{OptimalTauBisection} can be efficiently calculated by using the bisection method.
Based on Proposition \ref{Pro:OptimalPowerFast} and \ref{Prop:TauOptimal}, we can iteratively update $\bm{\chi}$ and $\tau$ until the objective function converges.
Given that the objective has converged, we project the obtained value of $\tau$ into $\mathcal{Z}_N$ to recover a integer-valued $N_t$. After that, we propose to once again update $\bm{\chi}$ using Proposition \ref{Pro:OptimalPowerFast} under constraint \eqref{KLBoundShortFurther} to refine the obtained solution.

\begin{remark}
\label{ComplexityESAO}
In this section,  the ES and the AO methods are proposed to maximize the ergodic sum-rate subject to a covertness constraint.
The ES method treats $\tau = \frac{N_d}{N}$ as a discrete variable and searches the optimal $\tau$ in $\mathcal{Z}_N$ exhaustively.
The AO method relaxes $\tau$ as a continuous variable in $(0,1)$ and optimizes $\tau$ and $\bm{\chi}$ alternately. Theoretically, the AO method only produces a sub-optimal solution.
Note that for each fixed value of $\tau$, both the ES and the AO methods need to compute the optimal $\bm{\chi}$, denoted by $\bm{\chi}^*(\tau)$, by using the bisection method to solve for the optimal dual variable according to Proposition \ref{Pro:OptimalPowerFast}.
Conceptually, the ES method needs to compute $\bm{\chi}^*(\tau)$ for $N-1$ times as $|\mathcal{Z}_N| = N-1$.
For the AO method, there is generally no guarantee on the maximal number of iterations before convergence.
However, our simulation results in Section \ref{Sec:Num2} show that only a few iterations are sufficient for the AO method to converge, and thus the AO method is computationally more efficient than the ES method.
\end{remark}

\section{Numerical Result}
We numerically evaluate the covert communication performance in this section.
We let the transmitter, the adversary, the jammer, and the $k$-th receiver ($1\leq k\leq K$) be in the same two-dimensional plane with their positions given by $\bm{s}_{\rm T}$, $\bm{s}_{\rm A}$, $\bm{s}_{\rm J}$, and $\bm{s}_{{\rm R}_k}$, respectively.
The distance-based path loss between node $a$ and node $b$ is $S_{a,b} = ||\bm{s}_a - \bm{s}_b||^{-4}$ where
$a,b\in\{{\rm T}, {\rm A}, {\rm J}, {\rm R}_1,  \cdots , {\rm R}_K\}$.
We set $\bm{s}_{\rm T} = (0,0)$, $\bm{s}_{\rm A} = (-d_{\rm A},0)$, and $\bm{s}_{\rm J} = (-d_{\rm J},0)$.
The positions of each receivers, $\bm{s}_{{\rm R}_k}$ for $1\leq k\leq K$, are independently and uniformly distributed in a circular region with the center and the radius being $(d_{\rm R},0)$ and $r_c$, respectively.
Unless specified, we set $d_{\rm A} = 150$ m, $d_{\rm J} = 250$ m, $d_{\rm R} = 150$ m, and $r_c = 30$ m,
$\sigma_{{\rm A}_1}^2 = \cdots =\sigma_{{\rm A}_K}^2 = -80$ dBm,
$\sigma_{{\rm R}_1}^2 = \cdots =\sigma_{{\rm R}_K}^2 = -80$ dBm, and
$\sigma_{\rm T}^2 = -80$ dBm, $P_{{\rm R}_1} = \cdots = P_{{\rm R}_K} = P_{\rm R} = 5$ dBm, and
$Q_{1} = \cdots = Q_{K} = Q = 25$ dBm.

\subsection{Covert communication over quasi-static channels}
\label{Sec:Num1}
In this subsection, we evaluate the covert communication performance under the condition that the channels are quasi-static.
Unless specified, we set $M = 20$ and $\epsilon = 0.005$.
\begin{figure}[t]
\begin{minipage}[t]{\linewidth}
\centering
\includegraphics[width=2.7 in]{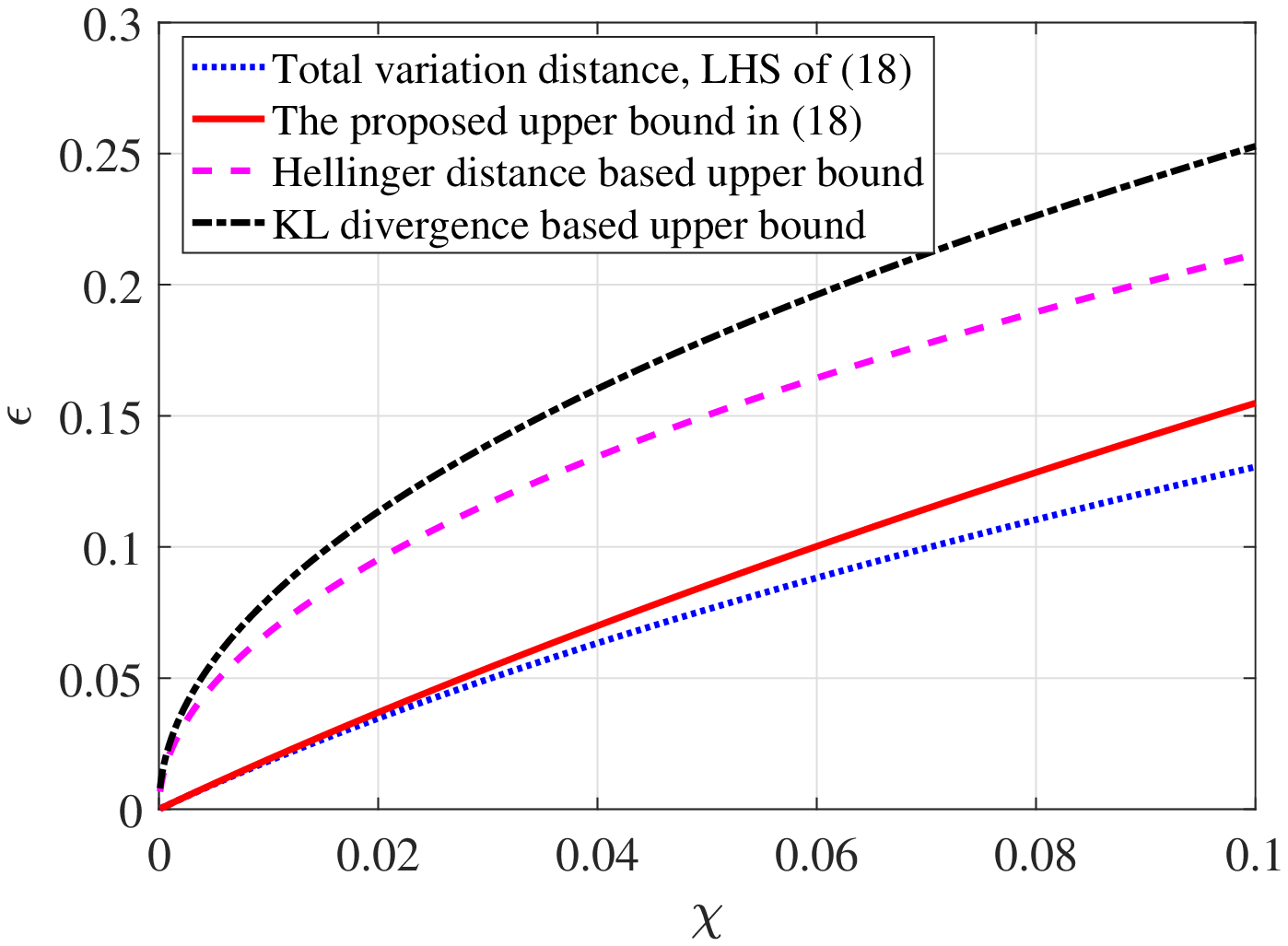}
\caption{$\mathbb{V}(f_U,f_V)$ and its upper bounds.}\label{Static1}
\end{minipage}%
\vspace{0.05 in}
\begin{minipage}[t]{\linewidth}
\centering
\includegraphics[width=2.7 in]{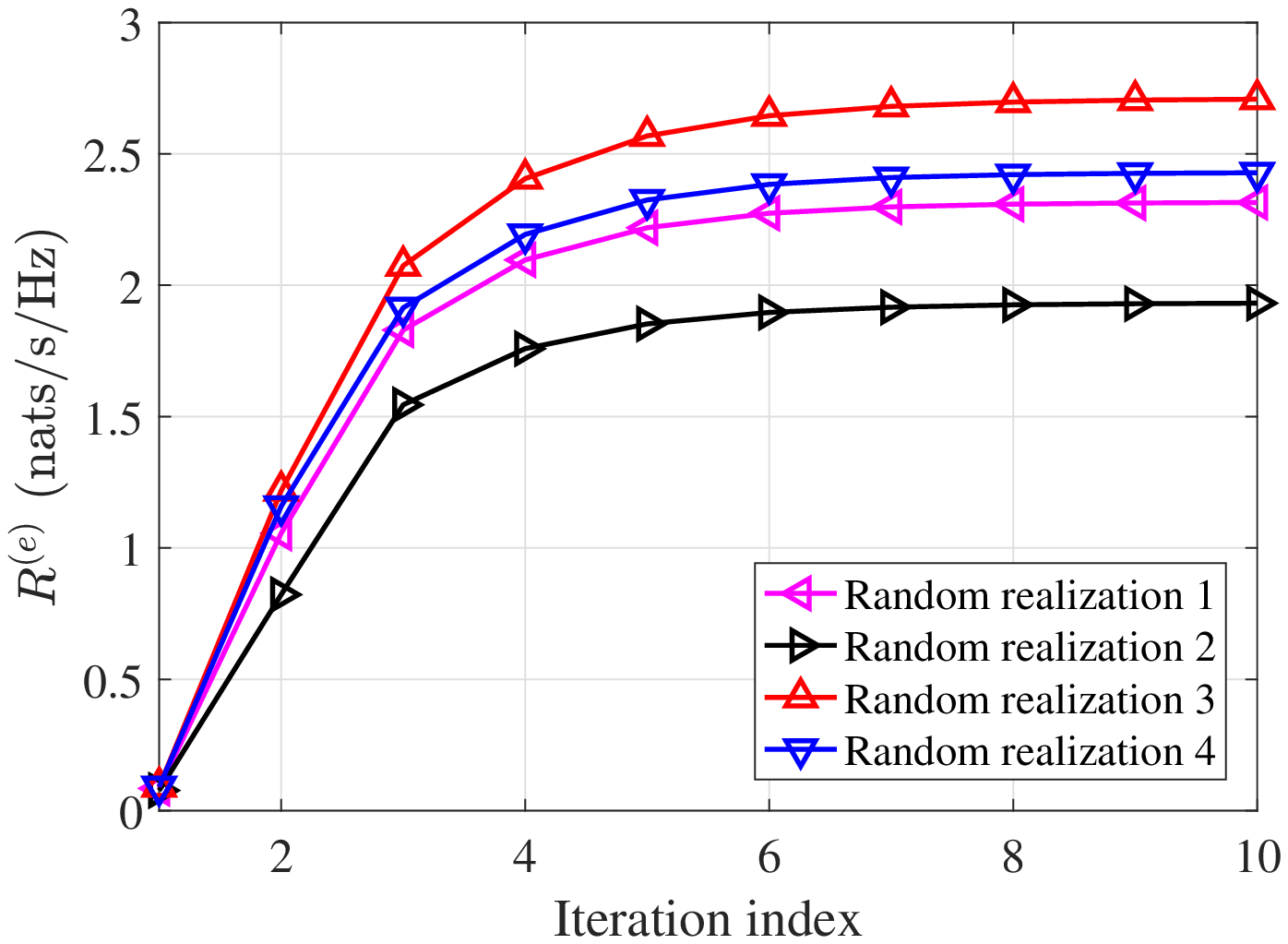}
\caption{Convergence of the SCA method with $K=3$.}\label{Static2}
\end{minipage}
\vspace{-6mm}
\end{figure}

In Fig. \ref{Static1}, we compare $\mathbb{V}(f_U,f_V)$ with its upper bound derived in \eqref{TVUpperBoundGeneralK}. Here, we set $K=2$ and $\chi_1 = \chi_2 = \chi$ for illustrative simplicity.
As there is no closed-form expression for $\mathbb{V}(f_U,f_V)$, we calculate it via numerical integration. Fig. \ref{Static1} reveals that the upper bound derived in \eqref{TVUpperBoundGeneralK} tightly approximates  $\mathbb{V}(f_U,f_V)$, especially when $\chi\rightarrow 0$. In Fig. \ref{Static1}, we also illustrate two widely used upper bounds on total variation distance, namely, the Hellinger distance based upper bound, see e.g.,
\cite[Theorem 13.1.2]{E.Lehmann}, and the KL-divergence based upper bound obtained by using Pinsker's inequality
\cite[Lemma 11.6.1]{Information}.
However,  Fig. \ref{Static1} shows that in the considered scenario, these bounds are not as tight as the proposed upper bound in \eqref{TVUpperBoundGeneralK}.

Fig. \ref{Static2} illustrates the convergence behavior of the SCA method.
We run the SCA method under four  groups of randomly generated system parameters.
Specifically, for each curve in Fig. \ref{Static2}, the locations of the receivers are randomly generated as introduced at the beginning of this section, and the channel coefficients are randomly generated by using the complex Gaussian distribution as assumed in Section II-A.
From Fig. \ref{Static2}, we can see that the SCA method converges within a few iterations.
In fact, through a large number of numerical experiments,
we observe that  under the considered system settings, the SCA method converges within $10$ iterations with the relative error smaller than $0.01$ in most cases.

\begin{figure}
\begin{minipage}[t]{\linewidth}
\centering
\includegraphics[width=2.7 in]{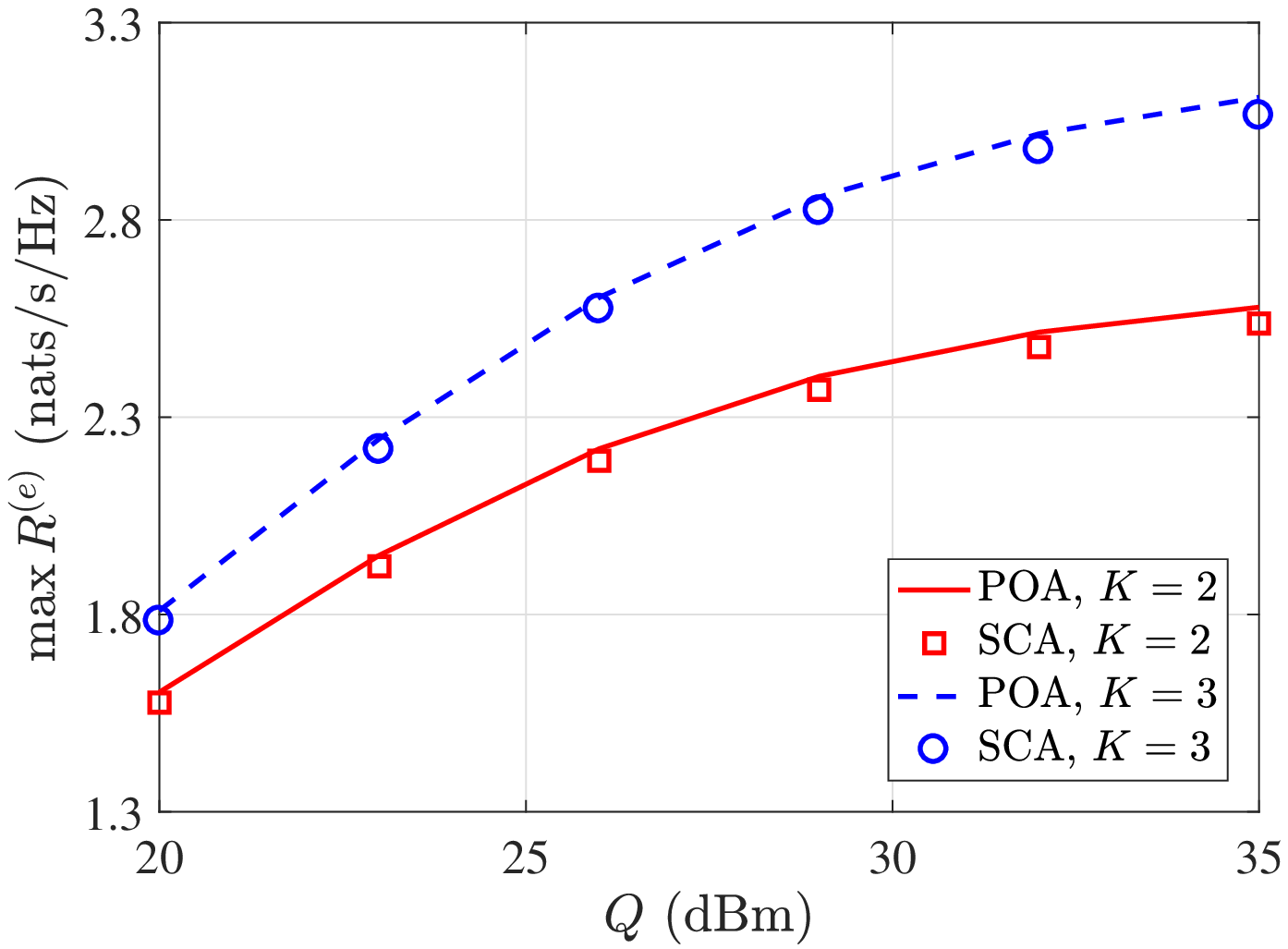}
\caption{$R^{(e)}$ versus $Q$.}\label{Static3}
\end{minipage}%
\vspace{0.05 in}
\begin{minipage}[t]{\linewidth}
\centering
\includegraphics[width=2.7 in]{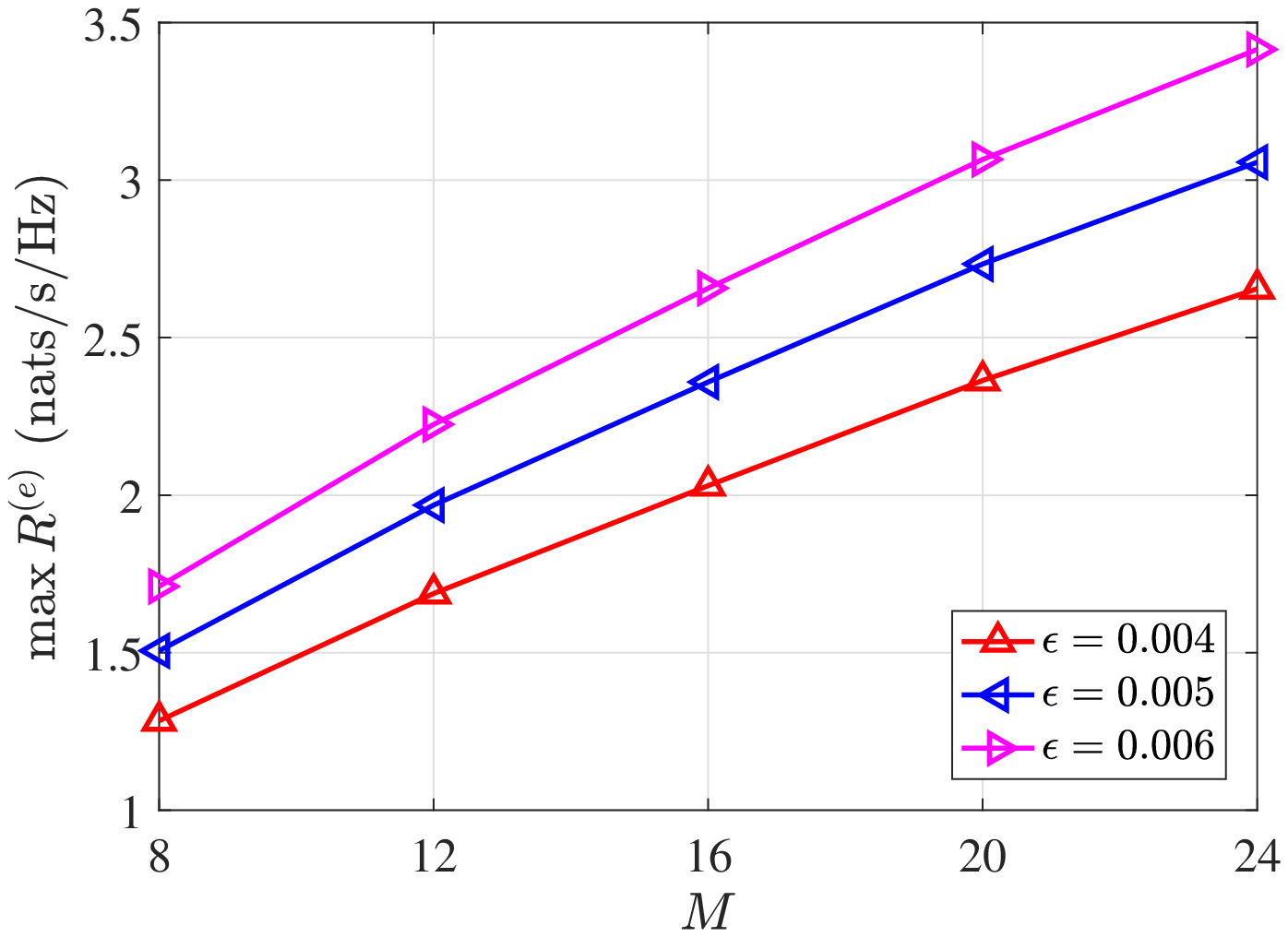}
\caption{$R^{(e)}$ versus $M$ with $K = 4$.}\label{Static4}
\end{minipage}
\vspace{-6mm}
\end{figure}

In Fig. \ref{Static3}, we plot the optimized effective sum-rate, $R^{(e)}$, versus the jamming power, $Q$, wherein the results obtained by using the SCA method are compared to those obtained by using the POA method.
Fig. \ref{Static3} indicates that the performance obtained by the SCA method is nearly optimal.
Besides, Fig. \ref{Static3} also shows that the covert communication performance gets improved as the jamming power increases.
This is because whether or not the covertness constraint \eqref{EffectiveRateAppx1Cons} is satisfied only depends on the values of $ \chi_1,\chi_2,\cdots,\chi_K$.
By definition, $\chi_k$ is proportional to $\frac{P_k}{Q_k}$, meaning that
we can proportionally increase $P_k$ with $Q_k$ without sacrificing the covertness.
And thus if  $P_1,P_2,\cdots,P_K$ are chosen such that $ \frac{P_1}{Q_1},\frac{P_2}{Q_2},\cdots,\frac{P_K}{Q_K}  $  are fixed constants and satisfy the covertness constraint, then it can be checked that the effective rate $r_k^{(e)}$ increases with $Q_k$ for $1\leq k\leq K$. Therefore, the optimal effective sum-rate also increases with the jamming power.

In Fig. \ref{Static4}, we plot the maximal effective sum-rate as a function of $M$.
Fig. \ref{Static4} inspires us that increasing $M$ is a promising method to enhance the covert communication performance.
In fact, in the considered scenario, with the knowledge about the channel state information (CSI), the transmitter can design beamforming vectors to enhance the SNRs at the receivers.
Note that due to the mismatch between the beamforming vector of the transmitter and the CSI of the adversary, increasing $M$ does not increase the signal power received by the adversary, meaning that the covert communication performance can be improved by using more antennas without sacrificing the covertness.

\subsection{Covert communication over fast-varying channels}
\label{Sec:Num2}
\begin{figure}[t]
\begin{minipage}[t]{\linewidth}
\centering
\includegraphics[width=2.7 in]{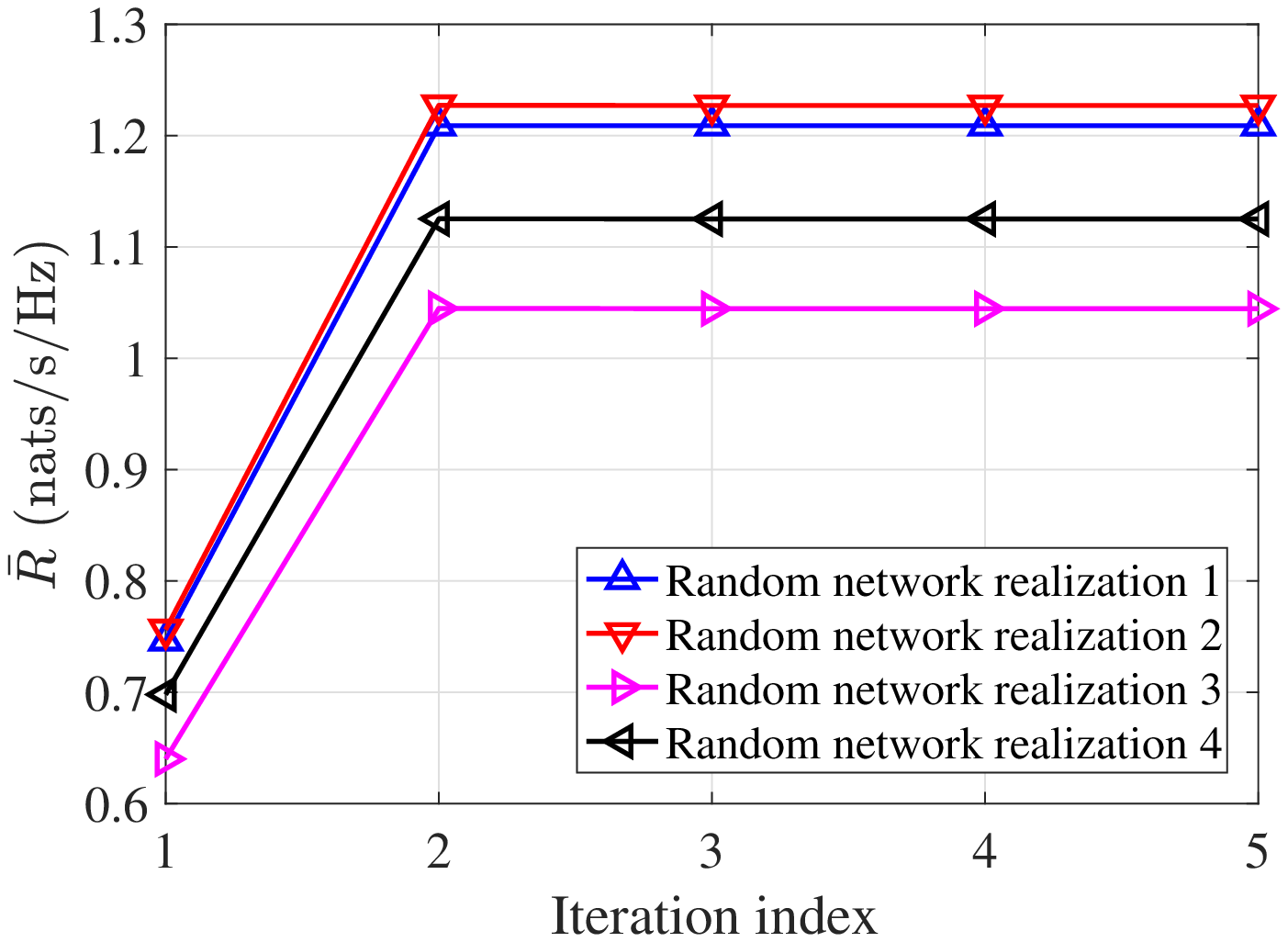}
\caption{Convergence of the AO method.}\label{Fast1}
\end{minipage}%
\vspace{0.05 in}
\begin{minipage}[t]{\linewidth}
\centering
\includegraphics[width=2.7 in]{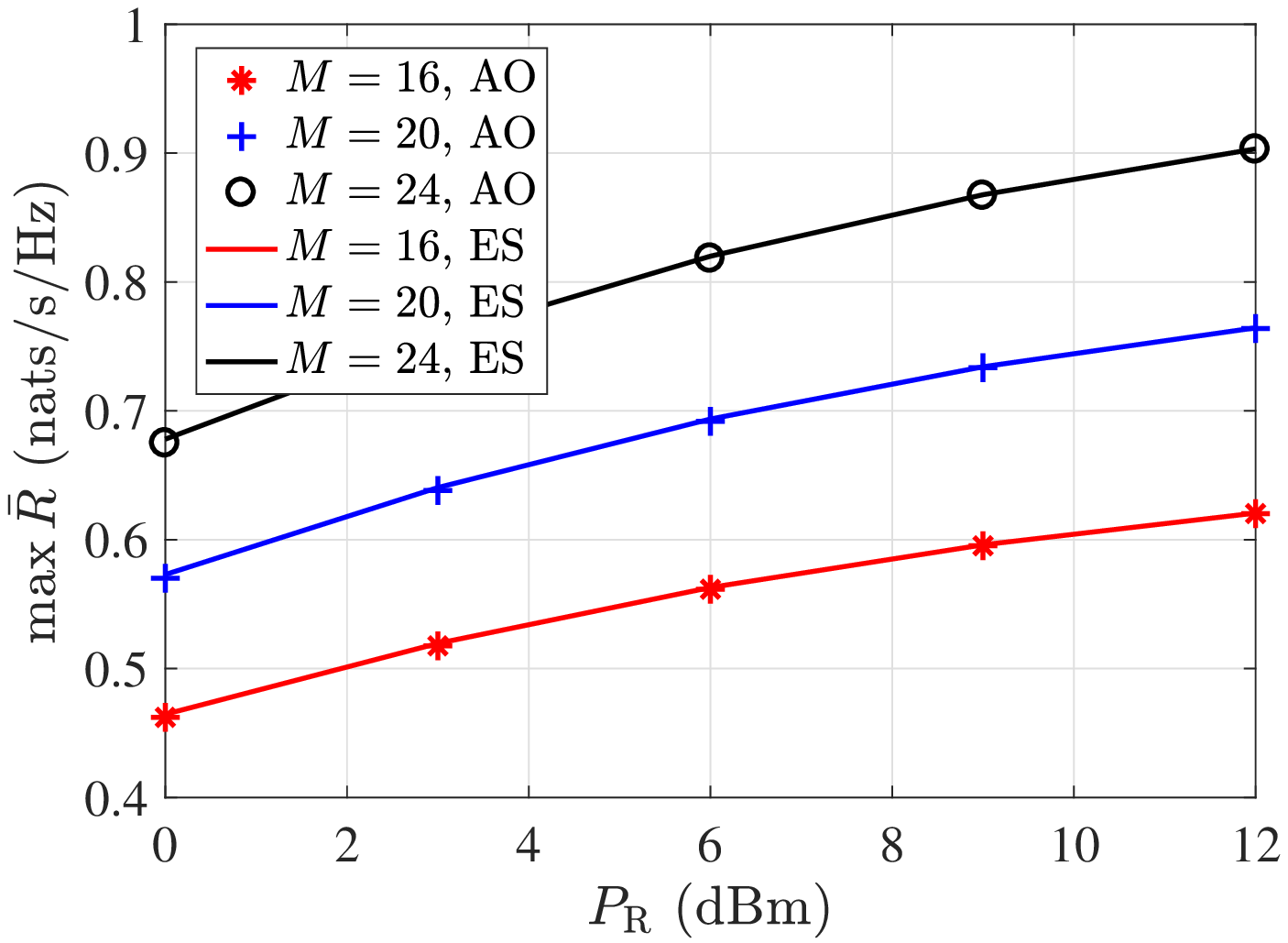}
\caption{$\bar{R}$ versus $P_{\rm R}$ with $\epsilon = 0.05$.}\label{Fast2}
\end{minipage}
\vspace{-6mm}
\end{figure}
In this subsection, we evaluate the ergodic sum-rate under the condition that the channels are fast-varying. Unless specified, we set $M = 20$, $N = 100$, $L = 100$, $\epsilon = 0.05$, and $K = 4$.

We first study the convergence behavior of the AO method introduced in Section \ref{Sec:AO}.
For each curve in Fig. \ref{Fast1}, we randomly generate the locations of the receivers and then apply the AO method to optimize the ergodic sum-rate.
We set the initial value of $\tau$ as $0.5$.
The value of $\bar{R}$ obtained in each iteration of the AO method are plotted.
As we can see from Fig. \ref{Fast1}, the AO method converges very fast. In fact, through a large number of numerical experiments,  we observe that the improvement on $\bar{R}$ becomes negligible after two iterations of the AO method in most cases.

In Fig. \ref{Fast2},  we show $\bar{R}$ as a function of $P_{\rm R}$. Both the AO and the ES methods are implemented to solve \eqref{Opt22}.
Note that the solutions obtained by the ES method are global optimal.
From Fig. \ref{Fast2}, it can be seen that the performance achieved by the AO method is nearly optimal.
Combining Fig. \ref{Fast1} and \ref{Fast2}, we conclude that the AO method is computationally efficient while causes little performance loss. Fig. \ref{Fast2} also reveals that the communication performance gets significantly improved as the $P_{\rm R}$ increases, which is due to that fact that with a larger value of $P_{\rm R}$, the transmitter can estimate the CSIs more accurately.

\begin{figure}[t]
\begin{minipage}[t]{\linewidth}
\centering
\includegraphics[width=2.7 in]{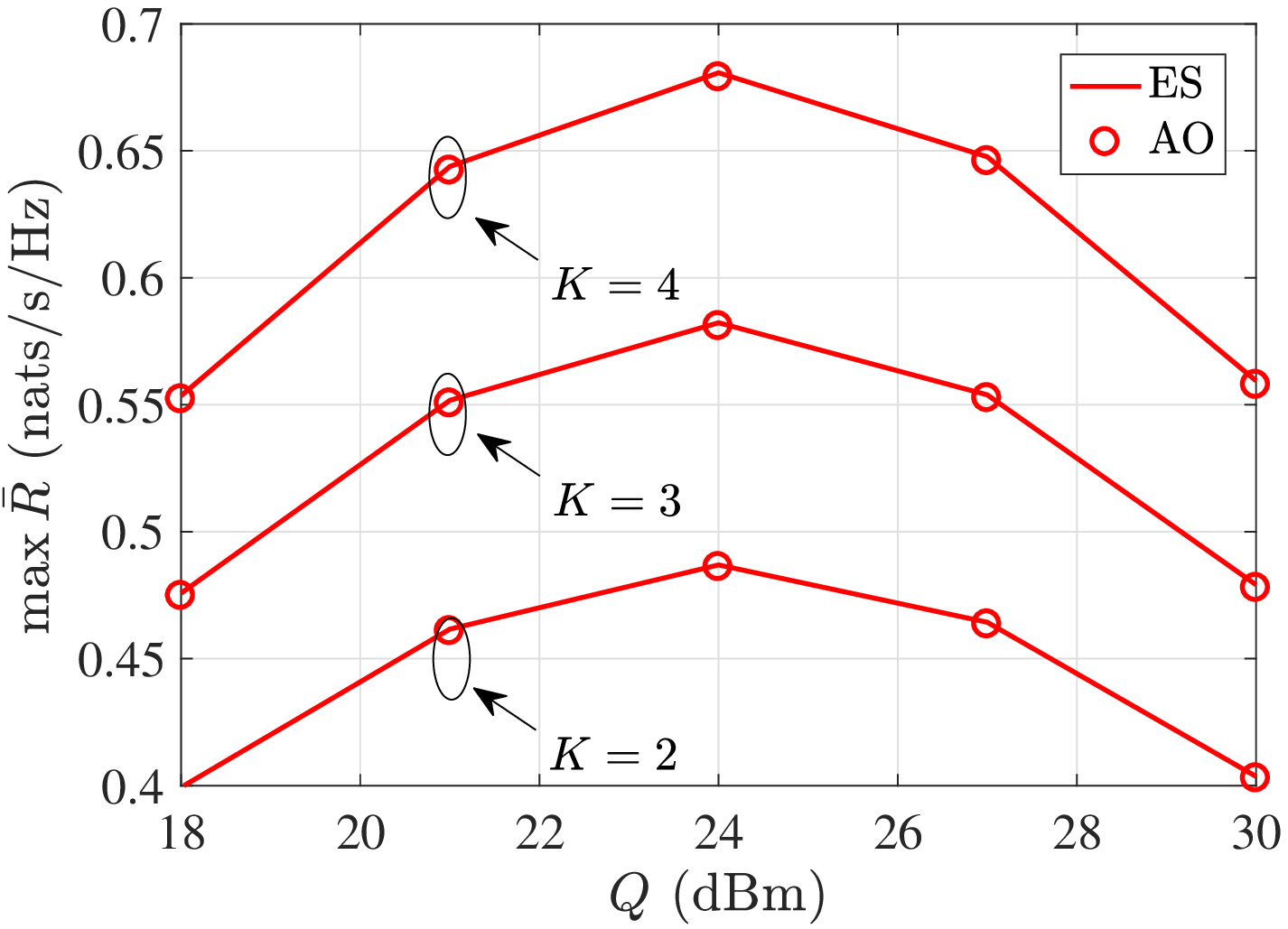}
\caption{$\bar{R}$ versus $Q$ with $K = 2,3,$ and $4$.}\label{Fast3}
\end{minipage}%
\vspace{0.05 in}
\begin{minipage}[t]{\linewidth}
\centering
\includegraphics[width=2.7 in]{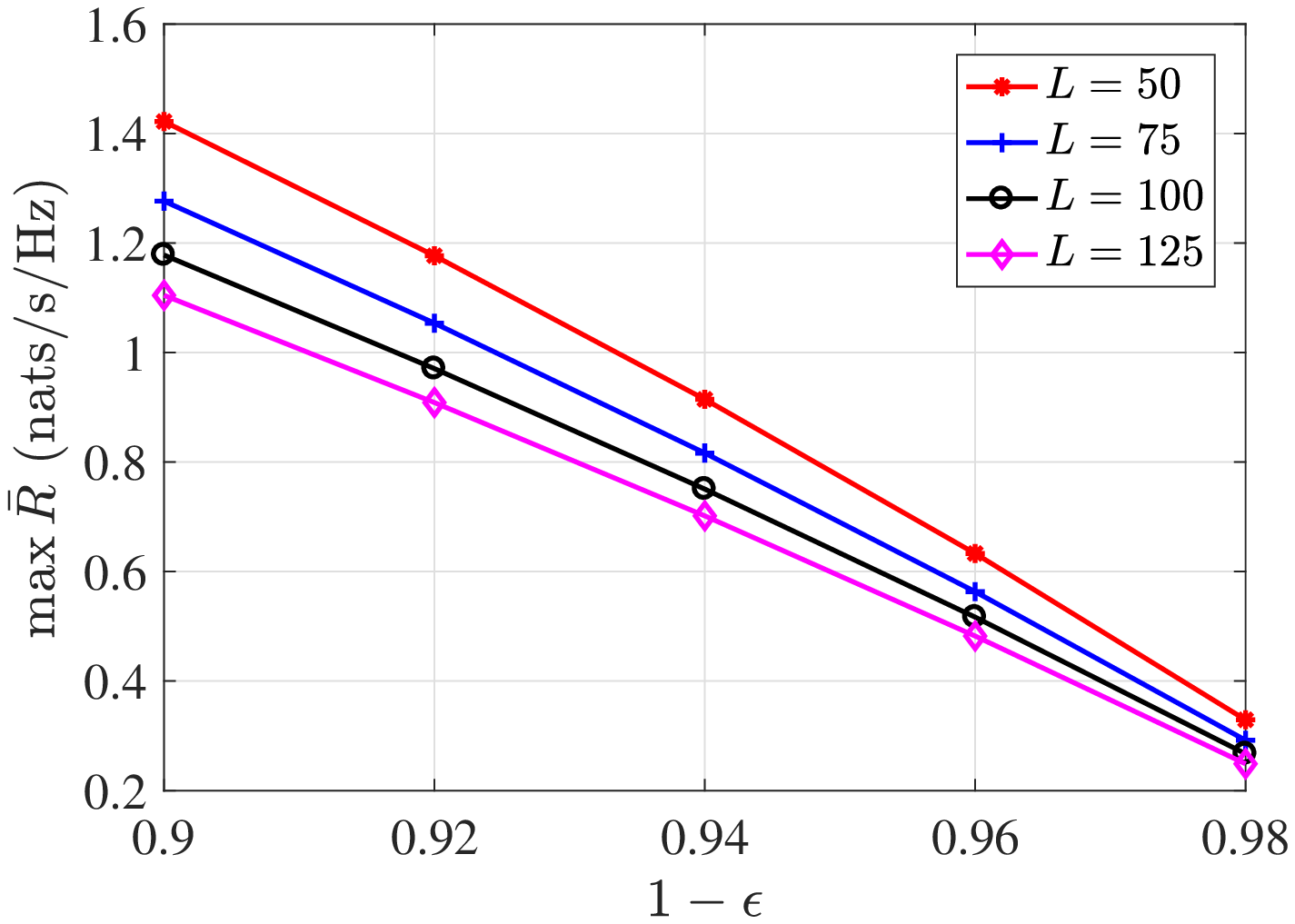}
\caption{$\bar{R}$ versus $1-\epsilon$ with $N\times L$ fixed as $1.5\times 10^3$.}\label{Fast4}
\end{minipage}
\vspace{-6mm}
\end{figure}

The influences of the jamming power on the ergodic sum-rate are plotted in Fig. \ref{Fast3}.
From Fig. \ref{Fast3}, we observe that increasing the jamming power does not always improve the communication performance, and there seems to exist an optimal value of the jamming power.
In fact, through numerical calculation, we observe that $\zeta(q_k,N - N_t)$ in constraint \eqref{KLBoundShortFurther} increases with $q_k$ (recall that $q_k \propto Q_k$). This means that subject to \eqref{KLBoundShortFurther}, the increase of $Q_k$ inevitably leads to the decrease of $\chi_k$, which finally reduces $\bar{R}$.
We notice that the effects of increasing the jamming power are quiet different in Fig. \ref{Static3} and \ref{Fast3}. Specifically, in Fig. \ref{Static3}, the covert communication performance improves as the jamming power increases, whereas in Fig. \ref{Fast3}, the covert communication performance becomes degraded when the jammer power is sufficiently large. This inspires us that in practice, the jamming power should be carefully designed according to the temporal dynamic properties of the wireless channel of the wireless channels.

In Fig. \ref{Fast4}, we illustrate the ergodic sum-rate, $\bar{R}$, as a function of the lower bound on the error probability of the adversary, i.e., $1-\epsilon$, wherein we fix $N\times L = 1.5\times 10^{3}$.
As $1-\epsilon$ approaches $1$, $\bar{R}$ monotonically decreases, which is the direct result of the reduced transmit power in order to satisfy the covertness constraint.
It is also worth noting that $\bar{R}$ decreases with $L$.
In fact, as $L$ increases, the adversary obtains more signal samples that are under different random channel realizations. This helps the adversary to average out the unknown fading channel coefficients, and thus leads to a more reliable detection result.

\section{Conclusion}
Covert communication between a transmitter and multiple receivers under the help of a friendly jammer has been studied.
When the channels are quasi-static, we maximized the sum of the effective rates.
When the channels are fast-varying, we maximized the sum of the ergodic rates. The optimal solutions to both of the two optimization problems can be numerically obtained. And we have also presented methods to search for sub-optimal solutions with relatively lower computational complexities.
Numerical results have been presented to show the covert communication performance.
We revealed that increasing the number of antenna at the transmitter is promising in improving the covert communication performance.
Besides, it has also been revealed that the existence of a friendly jamming greatly helps improving the covert communication performance.

\appendix
\subsection{The proof of Proposition 1}
\label{APP:Pro1}
Let $n\geq 1$ be an integer, by the definition of the total variation distance, we obtain that
\begin{align}
&2\mathbb{V}(f_1^{(n)} , f_0^{(n)}) = \int_{\mathcal{C}^{Kn}} \Bigg|f_1^{(n)}(\bm{W}) - f_0^{(n)}(\bm{W})  \Bigg|
\rmd \bm{W}  \nonumber \\
\overset{(a)}{=}& \int_{\mathcal{R}_{+}^{K}} \Bigg|
\prod_k \int_{\sigma_{{\rm A}_k}^2}^{\infty}\frac{z_k^{n-1}f_{U_k}(u_k)}{(n-1)!}\frac{\e^{-z_k/u_k}}{u_k^n} \rmd u_k \nonumber \\
&\quad -
\prod_k \int_{\sigma_{{\rm A}_k}^2}^{\infty}\frac{z_k^{n-1}f_{v_k}(v_k)}{(n-1)!}\frac{\e^{-z_k/v_k}}{v_k^n}  \rmd v_k \Bigg|
\rmd \bm{z} \nonumber \\
\overset{(b)}{=}&  \int_{\mathcal{R}_{+}^{K}}  \left| \prod_k f_{U_k}^{(n)}(y_k) - \prod_k f_{V_k}^{(n)}(y_k) \right| \rmd \bm{y} \nonumber \\
=&\left(\int_{\Omega} + \int_{\mathcal{R}_{+}^{K}\setminus\Omega }\right) \left|
\prod_k f_{U_k}^{(n)}(y_k) -
\prod_k f_{V_k}^{(n)}(y_k) \right|
\rmd \bm{y}\label{TVFnUfnV}
\end{align}
where $\bm{W} = [\bm{w}_1,\cdots,\bm{w}_K]$ with $\bm{w}_k = [w_{k,1},\cdots,w_{k,n}]^T$,
$f_1^{(n)}(\cdot)$ and $f_1^{(n)}(\cdot)$ are defined in \eqref{ProbabilityDensityFunction},
$f_{U_k}^{(n)}(x) \deq \int_{\sigma_{{\rm A}_k}^2}^{\infty}\frac{n^n x^{n-1}f_{U_k}(u)}{(n-1)!}\frac{\e^{-nx/u}}{u^n} \rmd u$,
$f_{V_k}^{(n)}(x) \deq \int_{\sigma_{{\rm A}_k}^2}^{\infty}\frac{n^n x^{n-1}f_{V_k}(v)}{(n-1)!}\frac{\e^{-nx/v}}{v^n} \rmd v$,
$\Omega \deq (0,G]^K$ for some sufficiently large positive constant $G$,
step $(a)$ is obtained by using step $(a)$, $(b)$, and $(c)$ in the derivation of \eqref{OriginalKLDiv},
and in step $(b)$, we make a change of variable $y_k \leftarrow z_k / n$ for $1\leq k \leq K$.
\begin{lemma}\label{Lem:1}
For $1\leq k\leq K$ and $x>0$, we have $\lim_{n\ri }f_{U_k}^{(n)}(x) = f_{U_k}(x)$ and $\lim_{n\ri }f_{V_k}^{(n)}(x) = f_{V_k}(x)$ almost everywhere. Besides, for $n \geq 2$, there exists some constraint $c>0$ (independent of $n$) such that for $\forall x>0$, $f_{U_k}^{(n)}(x)<c$ and $f_{V_k}^{(n)}(x)<c$.
\end{lemma}
\begin{lemma}\label{Lem:2}
$\mathbb{V}_2^{(n)}\deq\int_{\mathcal{R}_{+}^{K}\setminus\Omega }\left| \prod_k f_{U_k}^{(n)}(y_k) - \prod_k f_{V_k}^{(n)}(y_k) \right| \rmd \bm{y}$ $= o_G(1) + o_n(1)$.
\end{lemma}

Lemma \ref{Lem:1} indicates that for $\forall n\geq1$, $\left| \prod_k f_{U_k}^{(n)}(y_k) - \prod_k f_{V_k}^{(n)}(y_k) \right|$ is bounded above by some constant.
Therefore, by using the bounded convergence theorem, see e.g., \cite[Theorem 1.4]{M.Razaviyayn2014}, we have that
$\lim_{n\ri} \mathbb{V}_1^{(n)} = \int_{\Omega} |\prod_k f_{U_k}(y_k) -
\prod_k f_{V_k}(y_k) |  \rmd \bm{y} $ where $\mathbb{V}_1^{(n)} \deq \int_{\Omega}  |
\prod_k f_{U_k}^{(n)}(y_k) -
\prod_k f_{V_k}^{(n)}(y_k)  | \rmd \bm{y}$ for any fixed $G>0$.
Since $2\mathbb{V}(f_1^{(n)} , f_0^{(n)}) = \mathbb{V}_1^{(n)} + \mathbb{V}_2^{(n)}$, Proposition \ref{Pro:TVnri} is proved by using Lemma \ref{Lem:2} and letting $G\ri$.
\begin{IEEEproof}[Proof of Lemma \ref{Lem:1}]
For simplicity, we only prove for the case of $f_{V_k}^{(n)}(x)$, and the proof can be directly extended to the case of $f_{U_k}^{(n)}(x)$.
For $x\in(0,\sigma_{{\rm A}_k}^2)$, by definition, we have that
\begin{align}
f_{V_k}^{(n)}(x) & =
\frac{n}{x}\int_{\sigma_{{\rm A}_k}^2}^{\infty} f_{V_k}(v)
\frac{n^n x^{n}}{n!}\frac{\e^{-nx/v}}{v^n} \rmd v\nonumber \\
&\overset{(a)}{\leq}  \frac{\sqrt{ n}}{ \sqrt{2\pi}x }
\int_{\sigma_{{\rm A}_k}^2}^{\infty} f_{V_k}(v)
\mathrm{e}^{n\varrho(x/v)} \rmd v \\
&\overset{(b)}{\leq} \frac{\sqrt{ n}}{ \sqrt{2\pi}x }
\mathrm{e}^{n\varrho(x/\sigma_{{\rm A}_k}^2)  } \xrightarrow{n\ri} 0 = f_{V_k}(x),
\end{align}
where step $(a)$ is because $n!\geq \sqrt{2\pi n}n^n/\e^{-n}$ (Stirling's formula) and $\varrho(z) \deq 1 - z + \ln z$, and step $(b)$ follows from the fact that  $ \varrho(z) \leq \varrho(1) = 0$ for any $z>0$, and that $\varrho(z)$ is increasing in $z\in(0,1)$. For $x\in(\sigma_{{\rm A}_k}^2,\infty)$, we have that
$f_{V_k}^{(n)}(x)
= \left(\int_{\mathcal{X}_\xi} + \int_{\mathcal{X}_{\xi}^c}\right) \frac{f_{V_k}(v) n^n x^{n-1}}{v^n (n-1)!} \mathrm{e}^{-\frac{n x}{v}}\rmd v$,
where $\mathcal{X}_\xi \deq ((1-\xi)x,(1+\xi)x)$, $\mathcal{X}_\xi^{c} \deq (\sigma_{{\rm A}_k}^2,\infty)\setminus\mathcal{X}$, and $\xi$ is a small positive number such that $(1-\xi)x > \sigma_{{\rm A}_k}^2$.We have
\begin{align}
&\quad \int_{\mathcal{X}_\xi}\frac{f_{V_k}(v) n^n x^{n-1}}{v^n (n-1)!} \mathrm{e}^{-\frac{n x}{v}}\rmd v\nonumber \\
&= \frac{f_{V_k}(x^{\prime})n}{n-1} \int_{\frac{1}{1+\xi}}^{\frac{1}{1-\xi}}\frac{ n^{n-1} \hat{v}^{n-2}}{ (n-2)!} \mathrm{e}^{-n\hat{v}}\rmd \hat{v}
\xrightarrow{(a)}  f_{V_k}(x)\label{onesubset}\\
&\quad \int_{\mathcal{X}_\xi^c}\frac{f_{V_k}(v) n^n x^{n-1}}{v^n (n-1)!} \mathrm{e}^{-\frac{n x}{v}}\rmd v\nonumber \\
&\overset{(b)}{\leq}  \frac{1}{\hat{q}_k}\frac{n}{n-1} \int_{(0,\frac{1}{1+\xi})\cup(\frac{1}{1-\xi},\infty)} \frac{ n^{n-1} \hat{v}^{n-2}}{ (n-2)!} \mathrm{e}^{-n\hat{v}}\rmd \hat{v} \xrightarrow{n\ri} 0\label{nullsubset}
\end{align}
where $x^{\prime}$ is some real number that lies between $(1 - \xi)x$ and $(1 + \xi)x$, step $(a)$ is due to the fact that
$\int_{\frac{1}{1+\xi}}^{\frac{1}{1-\xi}}\frac{ n^{n-1} \hat{v}^{n-2}}{ (n-2)!} \mathrm{e}^{-n\hat{v}}\rmd \hat{v} \xrightarrow{n\ri} 1$ for any $\xi >0$ and thus obtained by first letting $n\ri$ and then $\epsilon \downarrow 0$, and
step $(b)$ is due to the fact that $f_{V_k}(x)\leq \frac{1}{\hat{q}_k}$ for $\forall x \geq \sigma_{{\rm A}_k}^2$.
In summary, for $\forall x\in(0,\sigma_{{\rm A}_k}^2)\cup(\sigma_{{\rm A}_k}^2,\infty)$, $\lim_{n\ri }f_{V_k}^{(n)}(x) = f_{V_k}(x)$.
Besides, by using the steps in \eqref{onesubset} and \eqref{nullsubset}, it can be shown that for $\forall x\geq 0$,
$f_{V_k}^{(n)}(x)  \leq \frac{1}{\hat{q}_k} \frac{n}{n-1} \leq \frac{2}{\hat{q}_k}$ and thus is bounded above by some constant that is independent of $n$.
\end{IEEEproof}
\begin{IEEEproof}[Proof of Lemma \ref{Lem:2}]
Define $\mathbb{V}_{2,V}^{(n)}\deq  \int_{\mathcal{R}_{+}^{K}\setminus\Omega } \prod_k f_{V_k}^{(n)}(y_k)  \rmd \bm{y}$ and $\mathbb{V}_{2,U}^{(n)} \deq \int_{\mathcal{R}_{+}^{K}\setminus\Omega }  \prod_k f_{U_k}^{(n)}(y_k) \rmd \bm{y}$.
It is straight that  $\mathbb{V}_2^{(n)} \leq \mathbb{V}_{2,V}^{(n)} + \mathbb{V}_{2,U}^{(n)}$. By the definition of $\Omega$, we have $ \mathbb{V}_{2,V}^{(n)} \leq \sum_{k=1}^K  \int_{G }^{\infty} f_{V_k}^{(n)}(y_k)  \rmd y_k$.
We further have that
$\int_{G }^{\infty} f_{V_k}^{(n)}(x) \rmd x =  \int_{G }^{\infty} \int_{\sigma_{{\rm A}_k}^2}^{\infty}\frac{n^n x^{n-1}f_{V_k}(v)}{(n-1)!}\frac{\e^{-nx/v}}{v^n} \rmd v  \rmd x = X_{k,1} + X_{k,2}$ where
$X_{k,1}\deq \int_{G }^{\infty}
\int_{\sigma_{{\rm A}_k}^2}^{\frac{G}{2}}\frac{n^n x^{n-1}f_{V_k}(v)}{(n-1)!}\frac{\e^{-nx/v}}{v^n} \rmd v  \rmd x $ and
$X_{k,2}\deq \int_{G }^{\infty}
\int_{\frac{G}{2}}^{\infty}\frac{n^n x^{n-1}f_{V_k}(v)}{(n-1)!}\frac{\e^{-nx/v}}{v^n} \rmd v  \rmd x $.
By changing the order of integration, $X_{k,1}$ satisfies
$X_{k,1} =  \int_{\sigma_{{\rm A}_k}^2}^{\frac{G}{2}} f_{V_k}(v) \Gamma(n, nG/v)  \rmd v \leq \Gamma(n, 2n ) = o_n(1)$.
For $X_{k,2}$, we have that $X_{k,2} \leq \int_{0}^{\infty}  \int_{\frac{G}{2}}^{\infty}\frac{n^n x^{n-1}f_{V_k}(v)}{(n-1)!}\frac{\e^{-nx/v}}{v^n} \rmd v  \rmd x
= \int_{\frac{G}{2}}^{\infty}f_{V_k}(v) \rmd v = o_G(1)$. Similarly, it can be shown that $\mathbb{V}_{2,U}^{(n)} = o_G(1)+ o_n(1)$, which completes the proof.
\end{IEEEproof}
\subsection{A brief introduction on the POA method}
\label{APP:POA}
\begin{figure*}[t]
\centering
\includegraphics[width=5  in]{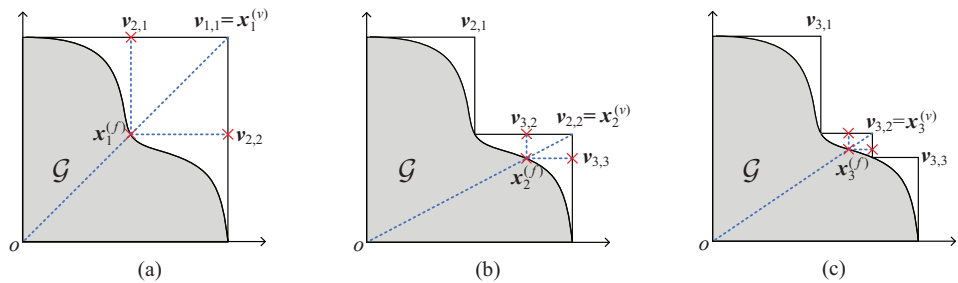}\\
\caption{Illustration of the POA method in a two-dimensional space, where the shadow area is the feasible set $\mathcal{G}$. Here, in the first iteration, $\mathcal{G}$ is approximated by $\mathcal{P}_1$ whose vertex set is $\mathcal{V}_1 = \{\bm{v}_{1,1}\}$, see Fig. \ref{POAIllusrate}(a). After the first iteration, we obtain an refined polyblock outer approximation, i.e., $\mathcal{P}_2$, whose vertex set if $\mathcal{V}_2 = \{ \bm{v}_{2,1}, \bm{v}_{2,2}\}$, see Fig. \ref{POAIllusrate}(b). Following similar steps, we obtain $\mathcal{P}_3$, whose vertex set is $\mathcal{V}_3 = \{ \bm{v}_{3,1}, \bm{v}_{3,2}, \bm{v}_{3,3}\}$, see Fig. \ref{POAIllusrate}(c). The iteration continues until some pre-defined termination criterion is satisfied.
}\label{POAIllusrate}
\vspace{-5mm}
\end{figure*}

In this part, we briefly introduce the POA method. We first present some basic definitions.
\begin{definition}[Monotonic functions]
\label{D:Increasing}
A function $h:\mathcal{R}^n\rightarrow \mathcal{R} $ is monotonically increasing if $h(\bm{x})\geq h(\bm{y})$ when $\bm{x}\geq \bm{y}$.
\end{definition}
\begin{definition}[Hyper-rectangles (a.k.a. Boxes)]
\label{D:Box}
Let $\bm{a},\bm{b}\in\mathcal{R}^n$ with $\bm{x}\leq \bm{y}$, then the set of $\bm{x}\in\mathcal{R}^n$ such that $\bm{a}\leq \bm{x}\leq \bm{b}$ is a box in $\mathcal{R}^n$, which is denoted by $[\bm{a},\bm{b}]$.
\end{definition}
\begin{definition}[Normal sets]
\label{D:NormalSet}
A set $\mathcal{X}$ is normal if $\forall \bm{x}\in\mathcal{X} $, then the box $[\bm{0},\bm{x}]\subseteq \mathcal{X}$.
\end{definition}
\begin{definition}[Polyblocks]
\label{D:PolyBlock}
A set $\mathcal{P}\in\mathcal{R}_+^n$ is a polyblock if  $\mathcal{P}= \cup_{\bm{v}\in\mathcal{V}}[\bm{0},\bm{v}]$, where
$\mathcal{V}$ is refer to as the vertex set of $\mathcal{P}$ with $|\mathcal{V}|<+\infty$.
\end{definition}
\begin{definition}[Monotonic optimization]
\label{D:MOP}
Let $h:\mathcal{R}^n\rightarrow \mathcal{R} $ be a monotonically increasing function, and $\mathcal{G}\subseteq [\bm{0},\bm{v}]$ for some $\bm{v}\in\mathcal{R}_+^n$ is a compact normal set with nonempty interior. Then,
$\max_{\bm{x}\in \mathcal{G}}~h(\bm{x})$ is a monotonic optimization problem.
\end{definition}

Based on the definitions above, let's consider a simple monotonic optimization problem in the form of
$\max_{\bm{x}\in \mathcal{G}}~h(\bm{x})$,
where $h:\mathcal{R}_+^n\rightarrow \mathcal{R}$ is continuous and monotonically increasing,
and $\mathcal{G} \subseteq \mathcal{R}_+^n$ is a compact normal set.
The basic idea behind the POA method to solve such a monotonic optimization problem is to iteratively generate a sequence of polyblocks to approximate the feasible set $\mathcal{G}$, i.e., $\mathcal{P}_1\supseteq \mathcal{P}_2 \supseteq  \cdots \supseteq \mathcal{P}_j \supseteq \cdots \supseteq \mathcal{G}$,
where $\mathcal{P}_j$ is the polyblock used to approximate $\mathcal{G}$ in the $j$-th iteration.
Denote by $\mathcal{V}_j$ the vertex set of $\mathcal{P}_j $. The $j$-th iteration of the POA method consists of the following steps:
\begin{enumerate}
\item \emph{Find the best vertex:} $\bm{x}_j^{(v)} = \mathop{\mathrm{argmax}}_{\bm{x}\in \mathcal{P}_j}~h(\bm{x}) =  \mathop{\mathrm{argmax}}_{\bm{x}\in \mathcal{V}_j}~h(\bm{x})$;
\item \emph{Project $\bm{x}_j^{(v)}$ into the feasible set $\mathcal{G}$:} $\bm{x}_j^{(f)} = \rho\bm{x}_j^{(v)}$, where $\rho =\mathop{\mathrm{max}}\{t: t\in (0,1), t \bm{x}_j^{(v)}\in \mathcal{G} \}$;
\item \emph{Updating the vertex set:} $\mathcal{V}_{j+1} = ( \mathcal{V}_{j} \setminus\bm{x}_j^{(v)} )\cup \{\bm{x}: \bm{x} = \bm{x}_j^{(v)} - \bm{E}_i( \bm{x}_j^{(v)}  - \bm{x}_j^{(f)} ), 1\leq i\leq n\}$, where $\bm{E}_i = {\rm diag}(\bm{e}_i)$ and $\bm{e}_i$ is the $i$-th standard basis of $\mathcal{R}^n$.
\end{enumerate}
The iteration continues until some certain termination criterion is satisfied, which we introduce later.
We illustrate the POA iteration introduced above in Fig. \ref{POAIllusrate} for better understanding.
As we can see that the POA method iteratively generates a sequence of feasible points.
Denote the set of the feasible points obtained after the $j$-th iteration as  $\mathcal{X}_j \deq \{x_1^{(f)}, x_2^{(f)},\cdots, x_j^{(f)} \}$, and among $\mathcal{X}_j$, denote the best feasible point as $\bm{x}_j^* = \max_{\bm{x}\in\mathcal{X}_j}~h(\bm{x})$.
Then, at the $(j+1)$-th iteration, it is reasonable to terminate the iteration process if $h(\bm{x}_{j+1}^{(v)}) - h(\bm{x}_j^*) \leq \delta$ where $\delta > 0$ a given tolerance. In fact, this means that $h(\bm{x}_j^*) \leq \max_{\bm{x}\in\mathcal{G}} h(\bm{x}) \leq \max_{\bm{x}\in\mathcal{P}_{j+1}} h(\bm{x}) = \max_{\bm{x}\in\mathcal{V}_{j+1}} h(\bm{x}) =  h(\bm{x}_{j+1}^{(v)}) \leq h(\bm{x}_j^*) + \delta $, and thus $\bm{x}_j^*$ is a $\delta$-optimal solution.

Note that in general, the number of points in the vertex set $\mathcal{V}_{j}$ increases exponentially with $j$, leading to a high computational complexity. There are some methods to reduce the points in $\mathcal{V}_{j}$, for example, \cite[Proposition 2.7]{Y.J.Zhang2012} and \cite[Section 3.3]{Y.J.Zhang2012}, which is helpful to reduce the computational burden.
We also note that in this part, we have only considered a simplified version of the canonical monotonic optimization formulation, which is, however, enough to handle the optimization problem in \eqref{EffectiveRateAppx2}.
For more details about the framework of monotonic optimization and the POA method, please refer to \cite{H.Tuy2005,Y.J.Zhang2012} and references therein.

\subsection{A brief introduction on the SCA method and the derivation of \eqref{SCAStep2}}
\label{APP:SCA}
Consider a general non-convex optimization problem $\min_{\bm{x}\in\mathcal{X} }\ g_0(\bm{x}) + h_0(\bm{x})$, where $\mathcal{X}\deq \{\bm{x}: g_t(\bm{x}) + h_t(\bm{x}) \leq 0, 1\leq t\leq T\}$ is the feasible set with $T$ being the number of the constraints, $ g_{t}(\bm{x})$ for $0\leq t\leq T$ are convex functions, and $h_{t}(\bm{x})$ for $0\leq t\leq T$ are non-convex functions.
The SCA method handles such a general non-convex optimization problem by replacing the non-convex function $h_t(\bm{x})$ with a convex approximation near some feasible point $\hat{\bm{x}}$, denoted by $\hat{h}_t(\bm{x};\hat{\bm{x}})$, and iteratively solving the resultant convex problem. Specifically, in the $j$-th iteration, the SCA method solves $\min_{\bm{x}\in\mathcal{X}_j }\ g_0(\bm{x}) + h_0(\bm{x};\bm{x}_{j-1})$ where $\mathcal{X}_j \deq \{\bm{x}: g_t(\bm{x}) + h_t(\bm{x};\bm{x}_{j-1}) \leq 0, 1\leq t\leq T\}$, where $\bm{x}_{j-1}$ is set to be the optimal solution obtained in the $(j-1)$-th iteration. Consequently, the SCA method generates a sequence of solutions $\{\bm{x}_1,\bm{x}_2,\cdots\}$. It has been shown in literature that for $\forall 0\leq t\leq T$, if $\hat{h}_t(\bm{x};\hat{\bm{x}})$ satisfies
1) $\hat{h}_t(\bm{x};\hat{\bm{x}})\leq h_t(\bm{x})$ for $\forall \bm{x}$,
2) $\hat{h}_t(\hat{\bm{x}};\hat{\bm{x}})= h_t(\hat{\bm{x}})$ for $\forall \bm{x}$,
3) $\frac{\partial h_t(\bm{x})}{\partial \bm{x}}\Big|_{\bm{x} = \hat{\bm{x}}} = \frac{\partial \hat{h}_t(\bm{x} ;\hat{\bm{x}})}{\partial \bm{x}}\Big|_{\bm{x} = \hat{\bm{x}}}$, and
4)$\hat{h}_t(\bm{x} ;\hat{\bm{x}})$ is continuous in $(\bm{x} ;\hat{\bm{x}})$,
then under some mild assumptions, the limiting point generated by the SCA method is a KKT solution to the original non-convex problem.
For more details about the SCA method, please see, e.g. \cite{M.Razaviyayn2014,A.Beck2010} and references therein.

In our problem in \eqref{SCAStep1},
the objective and the constraint in \eqref{SCAStep1C1} are non-convex.
To use the SCA method, in the $(j+1)$-th iteration, we approximate the non-convex part of the objective, i.e., $-||\bm{\alpha}+\bm{\beta}||^2$, by using its first order Taylor expansion.
For the non-convex constraint \eqref{SCAStep1C1}, we approximate it using $\frac{1}{2}\bm{t}^T\bm{\Lambda}_1^{(j)}\bm{t} + \frac{1}{2}\bm{\gamma}^T\bm{\Lambda}_2^{(j)}\bm{\gamma} \leq \epsilon$, where the definitions of $\bm{\Lambda}_1^{(j)}$ and $\bm{\Lambda}_2^{(j)}$ are presented below \eqref{SCAStep2}. It can be verified that the convex approximations adopted here satisfy the conditions mentioned above, and thus enable the SCA method.

\end{document}